\documentclass[a4paper,11pt]{article}
\pdfoutput=1 

\usepackage{jheppub,amsthm} 

\usepackage[utf8]{inputenc}
 
\theoremstyle{remark}
\newtheorem{remark}{Remark}[section]

\usepackage{booktabs,multirow, quotes}
\usepackage[normalem]{ulem}

\def\Hm{{\mathcal{H}}}

\def\Km{{\mathcal{K}}}
\def\Lm{{\mathcal{L}}}

\def\Om{{\mathcal{O}}}

\def\Vm{{\mathcal{V}}}

\renewcommand{\epsilon}{\varepsilon}
\renewcommand{\ge}{\geqslant}
\renewcommand{\le}{\leqslant}

\makeatletter
\def\@fpheader{\ }
\makeatother

\title{Scale without Conformal Invariance in Dipolar Ferromagnets}

\author[a]{Aleix Gimenez-Grau,}
\author[b]{Yu Nakayama,}
\author[a]{and Slava Rychkov}

\preprint{YITP-23-107}

\affiliation[a]{Institut des Hautes \'Etudes Scientifiques, 91440 Bures-sur-Yvette, France}
\affiliation[b]{Yukawa Institute for Theoretical Physics,
Kyoto University, Kitashirakawa Oiwakecho, Sakyo-ku, Kyoto 606-8502, Japan}

\emailAdd{gimenez@ihes.fr}
\emailAdd{yu.nakayama@yukawa.kyoto-u.ac.jp}
\emailAdd{slava@ihes.fr}

\abstract{We revisit critical phenomena in isotropic ferromagnets with strong dipolar interactions. The corresponding RG fixed point - dipolar fixed point - was first studied in 1973 by Aharony and Fisher. It is distinct from the Heisenberg fixed point, although the critical exponents are close. On the theoretical side, we discuss scale invariance without conformal invariance realized by this fixed point. We elucidate the non-renormalization of the virial current due to a shift symmetry, and show that the same mechanism is at work in all other known local fixed points which are scale but not conformal invariant. On the phenomenological side, we discuss the relative strength of dipolar and short-range interactions. In some materials, like the europium compounds, dipolar interactions are strong, and the critical behavior is dipolar. In others, like Fe or Ni, dipolar interactions are weaker, and the Heisenberg critical behavior in a range of temperatures is followed by the dipolar behavior closer to the critical point. Some of these effects have been seen experimentally.}
\begin{document} 

\maketitle

\flushbottom

\section{Introduction}

Most renormalization group fixed points of relevance to physics are
conformally invariant. Here we will describe an
experimentally observable continuous phase transition that is rotationally, translationally and scale invariant without being
conformal. This exceptional phase transition occurs in isotropic
ferromagnets possessing non-negligible dipolar interactions. Standard examples are the europium
compounds EuO and EuS. Aharony and Fisher pointed out in 1973
{\cite{FisherAharony,AharonyFisher,AF-PRBII}}\footnote{See also the review in \cite{Aharony:2023mqh}.} that the Curie point $T=T_c$ of such magnets
is not the usual Heisenberg fixed point, but a different one, possessing
slightly different values of critical exponents. The most dramatic difference
is that the longitudinal fluctuations of the order parameter are suppressed at
this new fixed point. This effect has been experimentally observed using
polarized neutron scattering {\cite{Koetzler}}. 

While most aspects of the physics of dipolar ferromagnets were well understood in the 1970s, the observation that the phase transition is scale without conformal appears to be new (it was first made in \cite{simons-talk}). We know of only one other such interacting experimentally relevant example, furnished by the theory of fluctuating membranes, as recently discussed in \cite{Mauri:2021ili}.
Other interacting field theories showing scale without conformal include gauge-fixed versions of gauge theories \cite{Nakayama:2016cyh}, which have not yet found experimental applications. Non-interacting examples of scale without conformal invariance include the theory of elasticity {\cite{Riva:2005gd}} and the Maxwell theory in 5d \cite{Jackiw:2011vz,El-Showk:2011xbs}. Holographic constructions of scale invariance without conformal invariance were studied in \cite{Nakayama:2009qu,
Nakayama:2012sn,Nakayama:2013is,
Nakayama:2016xzs,
Li:2018rgn}.

One issue that arises when discussing scale without conformal invariance in interacting models is the non-renormalization of the virial current - the vector operator $V_i$ whose divergence appears in the trace of the stress tensor: $T_{ii} =  - \partial_i V_i$. The scaling dimension of the virial current should be exactly $d-1$. It has been sometimes argued that this makes scale without conformal invariance unlikely, since non-conserved currents will generically pick up an anomalous dimension \cite{Rychkov:2016iqz,El-Showk:2011xbs}. How do dipolar magnets evade this? The reason is a shift symmetry acting on the field $U$, the Lagrange multiplier enforcing the transverse condition $\partial_i\phi_i=0$, where $\phi_i$ is the order parameter and at the same time the shift symmetry current. The virial current has the form $U\phi_i$, so that it transforms under the shift symmetry into the shift symmetry current. This relation implies that the scaling dimension of the virial current is protected from loop corrections due to interactions, although the shift symmetry current does get an anomalous dimension. This argument is one of our main new results, first presented in \cite{ERG-talk}.

Furthermore, we show that shift symmetry is also responsible for all previously known interacting theories realizing scale without conformal invariance.
A simple way to connect shift symmetry and lack of conformal invariance is as follows.
Say the shift symmetry acts on a local operator as $\Om(x) \to \Om(x) + c$, where $c$ is a constant.
Dimensional analysis implies that the scaling dimension of the shift current is $\Delta_J = d-1-\Delta_\Om$.
However, if $J_\mu$ is a primary operator, then conservation and conformal invariance would imply the shift current has dimension $\Delta_J = d-1$, leading to a contradiction.
Although there is a workaround for free theories, we conclude that a generic interacting theory can only be scale invariant.
Regarding the virial current, we show that there exists an operator of the schematic form $V_\mu \sim \Om J_\mu$ which has dimension exactly $\Delta_V = d-1$ with no loop corrections even in presence of interactions.
In the main text we present these arguments in more detail, and in particular we apply them to the membrane theory of Ref.~\cite{Mauri:2021ili}.
Ref.~\cite{Mauri:2021ili} previously proved the non-renormalization of the virial current in the membrane theory, via a different argument which also used a shift symmetry but in a less direct fashion.

In high-energy physics literature, the question of scale invariance vs conformal invariance is usually discussed for unitary theories \cite{Polchinski:1987dy,Dorigoni:2009ra,Dymarsky:2013pqa}. Everywhere in this paper we work in Euclidean signature, and unitarity is used synonymously with reflection positivity. From a wider field-theoretical perspective, unitarity is not \emph{in itself} a requirement for conformal invariance. Statistical physics furnishes many experimentally relevant critical theories which lack unitarity and yet are conformal, percolation and self-avoiding walks being two examples. As we will discuss, the dipolar fixed point, as well as all other fixed points realizing scale without conformal invariance with the help of the shift symmetry, are not unitary.

{\bf Outline.}
We start with Section \ref{sec:AF} where we introduce dipolar effects in the Heisenberg model, using several equivalent descriptions which are useful later in the paper. After that the reader may proceed along two independent routes:
\begin{enumerate}
	\item {\bf Pheno route.} Continue to Section \ref{sec:exp}, and the associated Appendices \ref{app:demag}-\ref{app:micro}, where we review experiments that found evidence for dipolar effects in critical ferromagnets, and provide a method to estimate which materials should exhibit important dipolar effects near the fixed point. We hope that this section will stimulate further computations of observable effects in this model, and their experimental studies.

\item {\bf Theory route.} Skip Section \ref{sec:exp} and go directly 
to Sections \ref{sec:scale} and \ref{sec:general}, and the associated Appendix \ref{sec:renorm}. In Section \ref{sec:scale} we argue that the dipolar fixed point is scale but not conformally invariant. We give two arguments - one from the two-point function $\langle \phi_i \phi_j\rangle$, and one from the trace of the stress tensor. The latter argument leads us to explain the non-renormalization of the virial current, which we do relying on the shift symmetry of the model.
We then show in Section \ref{sec:general} that the same arguments with minor modifications apply to all previously known examples of interacting scale but not conformal invariant fixed points.
\end{enumerate}

We conclude in Section \ref{sec:conclusion} with a summary and some possible future directions.

\section{Dipolar fixed point: Aharony-Fisher theory} 
\label{sec:AF}

Consider a three-dimensional (3D) isotropic ferromagnet in the vicinity of its
Curie temperature. Relation to actual materials will be discussed in Section \ref{sec:exp}. Here we write down the Landau-Ginzburg-Wilson (LGW) effective Hamiltonian
describing fluctuations of the three-component order parameter $\phi_i$, $i = 1, 2, 3$. The usual short-range Hamiltonian
is
\begin{equation}
  \int d^3 x \hspace{0.17em}\left( \frac{1}{2} \partial_i \phi_j \partial_i \phi_j 
 + \frac{1}{2} m^2  (\phi_i \phi_i) +\frac{\lambda}{4}
  (\phi_i \phi_i)^2\right) \hspace{0.17em} . \label{eq:LGW}
\end{equation}
This Hamiltonian gives rise to a renormalization group fixed point known as
the Heisenberg fixed point (or the Wilson-Fisher $O (3)$ fixed point). This fixed point can
be studied using the $\varepsilon$-expansion
{\cite{Wilson:1973jj}},\footnote{The literature being vast, more
references can be found in {\cite{Pelissetto:2000ek,Henriksson:2022rnm}}.} the numerical conformal bootstrap
{\cite{Kos:2013tga,Kos:2015mba,Kos:2016ysd,Chester:2020iyt}}, and Monte Carlo simulations (see e.g.~\cite{Hasenbusch-O3}).

However, any realistic 3D ferromagnet in addition to the short-range
interactions described by the above Hamiltonian will contain a long-range
dipolar interaction term
\begin{equation}
  V_{\text{dip}} = v \int d^3 x \int d^3 y \hspace{0.17em} U_{ij}  (x - y)
  \phi_i (x) \phi_j (y)\,, \label{eq:Vdip}
\end{equation}
where
\begin{equation}
  U_{ij} (x) = - \partial_{x_i} \partial_{x_j} \frac{1}{|x|} =
  \frac{\delta_{ij} - 3 \hat{x}_i  \hat{x}_j}{|x|^3} \hspace{0.17em} .
  \label{eq:U}
\end{equation}
In momentum space we have
\begin{equation}
	V_{\text{dip}} = 4 \pi v \int \frac{d^3 q}{(2 \pi)^3}  \hspace{0.17em}
	\frac{q_i q_j}{q^2} \phi_i (q) \phi_j (- q)\,. \label{eq:VdipQ}
\end{equation}

The term $V_{\text{dip}}$ appears because the field $\phi (x)$, proportional\footnote{In Section \ref{sec:exp}, we will change the normalization of the field $\phi$ so that it \emph{equals} the coarse-grained magnetization. Then the coefficient $v=1/2$. Here we find it more convenient to work in the normalization where the kinetic term $\frac{1}{2} \partial_i \phi_j \partial_i \phi_j$ is canonically normalized.} to
the coarse-grained magnetization at $x$, generates magnetic field throughout
the space, which in turn couples to the field $\phi (y)$. In Section \ref{sec:locdesc} below we show how \eqref{eq:Vdip} arises, with a positive coupling $v$, when integrating out the magnetic field. 

The symmetry of the LGW Hamiltonian \eqref{eq:LGW} was spatial $O (3)$ times
internal $O (3)$. The dipolar interaction breaks this symmetry to the diagonal
subgroup $O (3)$, under which $\phi_i$ transforms as a vector.\footnote{The mixing between space and internal symmetry is the reason we use Latin indices $i,j=1,\ldots,d$ for the dipolar model. For all other models, we use notation $\mu,\nu=1,\ldots,d$ for spacetime indices. }

As mentioned the term \eqref{eq:Vdip} will always be there for any material, because we can't turn off the Maxwell equations in an experiment (although we can do this in a Monte Carlo simulation). The best we can hope for is that the coupling $v$ is small. The actual size of coupling $v$ at the microscopic scale depends on the material and,
if $v$ is small, the Heisenberg fixed point could be a good description in a range of distances and reduced temperatures, see Section \ref{sec:exp}. But at sufficiently long distances and sufficiently close to the critical point, which may or may not be experimentally resolvable in practice, the pure Heisenberg description breaks down and $v$ needs to be taken into account. 

Indeed the term $V_{\text{dip}}$ is strongly relevant - it
has the same dimension as the mass term in \eqref{eq:LGW}. Moreover, it is the only nonlocal term in the Hamiltonian, so it will not get direct
renormalization contributions from local couplings.\footnote{There will be
indirect renormalization effect due to wavefunction renormalization, which is
small because the anomalous dimension of $\phi$ is small.} In the deep infrared (IR), the
effective coupling $v$ will grow to $+ \infty$. The effect of this will be
quite dramatic: at the IR fixed point all longitudinal fluctuations of the
order parameter will be suppressed
\begin{equation}
  \partial_i \phi_i = 0 \qquad \text{(at the IR fixed point)} \hspace{0.17em}
  . \label{eq:long}
\end{equation}
(Note that the integrand in \eqref{eq:VdipQ} can be written as $| q_i \phi_i
(q) |^2 / q^2$.) The fixed point for the remaining transverse fluctuations can
still be studied in the $\varepsilon$-expansion, but the usual procedure
should be modified. If one is interested in fixed-point physics only, one can
study the renormalization group (RG) flow of LGW Hamiltonian \eqref{eq:LGW} imposing the constraint
\eqref{eq:long} all along the flow. This means that one is working with the
transverse propagator
\begin{equation}
  \langle \phi_i (q) \phi_j (- q) \rangle = \frac{\delta_{ij} - q_i q_j /
  q^2}{q^2 + m^2}
\end{equation}
instead of the usual scalar field propagator. The propagator remains transverse along the RG flow. One works in $d = 4 - \varepsilon$ dimensions, in
which case $\phi_i$ is a $d$-component order parameter. As usual, one computes
the beta function for the quartic coupling, but since the diagrams are
computed with a different propagator, the actual beta-function coefficients are
a bit different. The quartic coupling flows to an IR fixed point, which we call "dipolar" (it was called
``isotropic dipolar'' in {\cite{AharonyFisher}}). The
anomalous dimensions of operators $\phi_i$, $\phi^2$,
$(\phi^2)^2$, etc, and the usual critical exponents can be computed as
power series in $\varepsilon$. These computations have been carried out in \cite{AharonyFisher,AF-PRBII} to
order $\varepsilon^2$, see Table II in {\cite{DombGreenVol6}}, p.~388.
Although these series are different from those for the Heisenberg case, the
numerical values extrapolated to $\varepsilon = 1$ come out
close. Recently, critical exponents were computed at three loops
directly in $d = 3$ {\cite{Kudlis:2022rmt}}, confirming the closeness to the
Heisenberg values.

The critical exponents being close to Heisenberg ($H$), the most dramatic feature of the
dipolar ($D$) fixed point remains the suppression \eqref{eq:long} of the
longitudinal fluctuations of the order parameter. The critical two-point (2pt)
function of the order parameter takes the form:
\begin{eqnarray}
  \langle \phi_i (q) \phi_j (- q) \rangle & = & \frac{\delta_{ij} - q_i q_j /
  q^2}{|q|^{2 - \eta_{D}}},
\label{2ptmom}
\end{eqnarray}
to be compared with $\langle \phi_i (q) \phi_j (- q) \rangle = \delta_{i j} /
|q|^{2 - \eta_{H}}$ at the Heisenberg fixed point. This suppression has been
seen with polarized neutron scattering, see Section \ref{sec:exp}. As we will see in Section \ref{sec:scale}, the 2pt function is compatible with conformal invariance at the Heisenberg but not at the dipolar fixed point.

\begin{remark}
	In this paper, we will be neglecting the cubic perturbation $\sum_{i=1}^3 \phi_i^4$ of the Hamiltonian \eqref{eq:LGW}. Such a perturbation, which may arise due to spin-orbit coupling, is usually assumed to be small in ferromagnets. It is known that this perturbation is (very weakly) relevant at the 3D Heisenberg fixed point \cite[Sec.~11.3]{Pelissetto:2000ek},
	\cite{Chester:2020iyt,Hasenbusch:2022zur}.
	\end{remark}

\subsection{Manifestly local description}
\label{sec:locdesc}

In the previous discussion, we perturbed the LGW Hamiltonian by a non-local dipolar interaction, and we argued that the infrared behavior is local, because dipole interactions simply impose the local constraint \eqref{eq:long}.
In this section, we reach the same conclusion starting from a manifestly local Hamiltonian.

As anticipated below equation \eqref{eq:Vdip}, the field $\phi_i(x)$ is proportional to the coarse-grained magnetization, and as such, it couples to the dynamical magnetic field $B_i$.
The precise form of the coupling is
  \begin{align} 
  	\int d^d x \left (\frac{1}{2} \partial_i \phi_j \partial_i \phi_j - z B_i \phi_i + \frac{B_i^2}{8\pi}  + \frac{m_0^2}{2} (\phi_i \phi_i)  +\frac{\lambda}{4} (\phi_i \phi_i)^2  \right)\,, 
  	\label{eq:HB}
  \end{align}
where $B_i = (\nabla \times A)_i$ is the magnetic field. The coefficient $z$ is the proportionality factor in $\phi_i= z M_i$ where $M_i$ is the coarse-grained magnetization (we will set $z=1$ in Section \ref{sec:exp} and Appendix \ref{app:demag}).
Because the vector potential $A_i$ appears quadratically, we can solve its equations of motion (EOM) exactly and plug them back into the action.
The result is the sum of \eqref{eq:LGW} and \eqref{eq:Vdip}, upon identifying
\begin{equation}
 \label{shift}
 m^2=m_0^2-4\pi z^{2}\, , \qquad
 v = \frac{z^2}{2} \, .
\end{equation}

The Hamiltonian \eqref{eq:HB} can also be used to show the suppression of longitudinal fluctuations.
We rewrite \eqref{eq:HB} by imposing the Bianchi identity via a Lagrange multiplier $U$:
  \begin{align}
  	\int d^d x \left (\frac{1}{2} \partial_i \phi_j \partial_i \phi_j  - z B_i \phi_i + \frac{B_i^2}{8\pi} + \frac{m_0^2}{2} (\phi_i \phi_i)  + \frac{\lambda}{4} (\phi_i \phi_i)^2  - U \partial_i B_i \right)\,.
  	\end{align}
  Note that the field $U$ coincides, up to rescaling, with the magnetic potential called $U$ in App.~\ref{app:demag}. Integrating out the unconstrained $B_i$ we obtain
  \begin{align}
 \int d^d x \left (\frac{1}{2} \partial_i \phi_j \partial_i \phi_j  - 2\pi (\partial_i U - z \phi_i)^2 + \frac{m_0^2}{2} (\phi_i \phi_i) + \frac{\lambda}{4} (\phi_i \phi_i)^2  \right) .
  \end{align}
Dropping the term $(\partial_i U)^2$, irrelevant in the long-wavelength limit, we obtain the effective Hamiltonian
\begin{align}
    \label{effLagr}
    \mathcal{H} = \int d^d x \left (\frac{1}{2} \partial_i \phi_j \partial_i \phi_j   - U \partial_i \phi_i + \frac{m^2}{2} (\phi_i \phi_i)  + \frac{\lambda}{4} (\phi_i \phi_i)^2 \right) \, ,
  \end{align}
up to a rescaling of $U$.
This shows that the role of $U$ is to impose the local constraint \eqref{eq:long}.

The scale invariant dipolar fixed point is obtained by fine-tuning $m^2$ at the critical value. From now on we work with the fine-tuned mass, which in dimensional regularization corresponds to $m^2 = 0$. The RG flow of $\lambda$ is attractive toward the fixed point value of $\lambda_*$.

Let us briefly compare theory \eqref{effLagr} to theory \eqref{eq:LGW} with constraint \eqref{eq:long} imposed ``by hand''. Correlation functions of $\phi_i$ are the same in both theories. In theory \eqref{effLagr} we have an additional local field $U$. It should not be too surprising that such an additional local field could be added to the theory. In fact, the original microscopic theory had the magnetic field $B_i$, and correlation functions of $U$ can be traced back to the (long wavelength limit of) correlators of $B_i$. The non-interacting 2pt functions of $U$ with itself and with $\phi_i$ are given by
\begin{equation}
	\langle U(q) U(-q)\rangle_0 = -1\,\qquad 	\langle U(q) \phi_i(-q)\rangle_0 = i q_i/q^2\,.
  \label{eq:freeprops}
\end{equation}
In perturbation theory, the field $U$ appears only at the external legs since there are no vertices involving it. Because the propagator $\langle \phi_i\phi_j\rangle_0$ is transverse, the 2pt function $\langle U \phi_i\rangle$ is not renormalized. So the anomalous dimension of $U$ will be the opposite to that of $\phi_i$:
\begin{equation}
	\label{phiU}
	\Delta_\phi = (d-2)/2+\gamma_\phi,\qquad \Delta_U = d/2-\gamma_\phi\,.
	\end{equation}
This is easy to understand: the wavefunction renormalization of $\phi_i$ and of $U$ comes from the $\phi_i$ self-energy $\Pi_{ij}$ (the sum of 1PI irreducible diagrams). In the $\phi_i$ case, $\Pi_{ij}$ is iterated, while for $U$ only the linear in $\Pi_{ij}$ term contributes, due to the transversality of $\langle \phi_i\phi_j\rangle_0$.

\section[Phenomenology and experiments]{Phenomenology and experiments\protect\footnote{Readers interested primarily in scale without conformal may proceed directly to Section \ref{sec:scale}.}}
\label{sec:exp}

Experimentally, dipolar behavior has been reported in some ferromagnets
(EuO, EuS), while in others, like Ni, Fe, it is harder to see, and in fact
they are usually assumed to exhibit Heisenberg behavior. Here we would like to
discuss why this is so, and how can one guess a priori which behavior to
expect from a given material, depending on the range of temperatures and
distance lengths used to probe the system.

For this discussion, it helps to normalize the field $\phi$ so that it equals
the (coarse-grained) microscopic magnetization: 
\begin{equation}
	\phi_i = M_i \, . \label{normM}
\end{equation}
The finite-temperature partition function is given by
\begin{equation}
	Z = \int D \phi\, e^{- \beta \mathcal{H} [\phi]} \, ,
\end{equation}
where the Hamiltonian, including the dipolar term, is given by:
\begin{equation}
	\mathcal{H} [\phi] = \int d^3 x \left( \frac{1}{2} a
	(\partial_i  \phi_j)^2 + \frac{1}{2} b \phi_i^{2} \right) + \frac{1}{2}  \int d^3 x \int d^3 y \; U_{ij}  (x
	- y) \phi_i (x) \phi_j (y) \, . \label{Hpheno}
\end{equation}
We are omitting here the quartic interaction term which will not play a role
in the present discussion. Notably, in the chosen normalization of $\phi$, the
coefficient of the dipolar term is completely
fixed.\footnote{\label{note:int-out}We obtain this term by integrating out $B_i$ from \eqref{eq:HB} with $z=1$, see Eq.~\eqref{shift}. See also App. \ref{app:demag}.} In presence of an external magnetic field $B^{(0)}$, the Hamiltonian should be perturbed by $-B^{(0)}_i \phi_i$, whose normalization is also fixed. This is important for discussing susceptibility measurements.

The inverse propagator of $\phi$ is given, up to overall rescaling, by
\begin{equation}
	G^{- 1}_{i j} (q)
	\propto (q^2 + \xi^{- 2}) \delta_{i j}
	+ q_d^2 \, \frac{q_i q_j}{q^2} \, , \label{Ginv}
\end{equation}
where we defined two important quantities:
\begin{equation}
	\xi = (a / b)^{1 / 2} \, , \qquad
	q_d = (4 \pi / a)^{1 / 2} \, . \label{xi}
\end{equation}
The $\xi$ is the correlation length, which goes to infinity when $b
\rightarrow 0$, as the critical point is approached. The $q_d$ is the dipolar
wavevector, which determines the range where the dipolar effects become
important. The propagator, obtained by inverting \eqref{Ginv}, is given by
\begin{equation}
	G_{i j} (q)
	\propto \frac{1}{q^2 + \xi^{- 2}} \left( \delta_{i j} -
	\frac{q_i q_j}{q^2} \right)
	+ \frac{1}{q^2 + \xi^{- 2} + q_d^2}  \frac{q_i q_j}{q^2} \, .
\end{equation}
We thus have two regimes, distinguishing between the short-range (Heisenberg)
and the dipolar behavior:
\begin{eqnarray}
	\text{Short range:} & \;
	\xi^{- 2} + q^2 \gg q_d^2 & \quad \Rightarrow \quad
	G_{i j}(q) \propto \frac{\delta _{i j}}{q^2 + \xi^{- 2}} \, , \\
	\text{Dipolar:} &  \;
	\xi^{- 2} + q^2 \ll q_d^2 & \quad \Rightarrow \quad
	G_{i j} (q) \propto \frac{1}{q^2 + \xi^{- 2}}
	\left( \delta_{i j} - \frac{q_i q_j}{q^2} \right) \, .
\end{eqnarray}
It is only in the second regime, where the propagator (which can be studied
e.g. using polarized neutron scattering) will show longitudinal suppression.

Thus, we see that to access experimentally the dipolar regime, two conditions
have to be satisfied. First, the correlation length $\xi$ must be sufficiently
large: $\xi^{- 1} \ll q _d$. This, according to \eqref{xi}, translates into $b
\ll 4 \pi$. In other words, we must be sufficiently close to the critical
point located at $b = 0$. In addition, the scattered neutrons have to be
sufficiently soft: $q \ll q_d$.

Let us focus on the criterion $b \ll 4 \pi$. To be useful, this criterion has
to be translated as a constraint on the reduced temperature $t = (T - T_c) /
T_c$.

For a given material, we can determine constants $a, b$ in the LGW
Hamiltonian doing experiments far away from $T_c$, when neglect of the quartic
interaction in \eqref{Hpheno} is justified. The constant $b$ can be determined by
measuring magnetization in the applied external uniform magnetic field
$B^{(0)}$ at $t = (T - T_c) / T_c = O (1)$ (i.e. far away from the transition
point) and using the relation
\begin{equation}
	\phi_i = \frac{B_i^{(0)}}{b + 4 \pi D_i} \, , \label{eq:demag}
\end{equation}
where $D_i$ is the demagnetizing factor, depending on the shape of the sample
(see App. \ref{app:demag}). The constant $a$ can be determined measuring the
correlation length $\xi$ (e.g. via neutron scattering) and using \eqref{xi}.

Once we determine $b$ at $t = O (1)$, we can extrapolate it to $t \ll 1$. In
the Gaussian approximation as in \eqref{Hpheno}, $b$ would be proportional to
$t$. In the critical region it is more appropriate to use the relation
corrected for the presence of critical exponents:
\begin{equation}
	b = C^{- 1} t^{\gamma}, \label{Bt}
\end{equation}
where $\gamma \approx 1.4$ is the Heisenberg susceptibility exponent (see Eq.~\eqref{eq:gammaHeis}) and $C$
is a dimensionless constant. Let us define $t_d$ to be the temperature such that $b(t_d) = 4 \pi$, i.e.
\begin{equation}
	t_d = (4 \pi C)^{1 / \gamma} .
\end{equation}
Then the dipolar behavior may be seen for $| t | \ll t_d$ while for larger $t$ we expect to see the Heisenberg behavior.
The same criterion to determine $t_d$ was proposed in \cite{PhysRevLett.51.833}.

Similarly, $\xi$ depends on $t$ according to:
\begin{equation}
	\xi = (a / b)^{1 / 2} = f^+ t^{- \nu}, \label{xit}
\end{equation}
where $f^+$ is a constant and $\nu \approx 0.7$ is the correlation length
exponent. Eqs. \eqref{Bt} and \eqref{xit} are consistent if $\gamma = 2 \nu$,
which is not exactly true, but is approximately true because $\eta$ is small.
Such an approximation is acceptable here, as we are aiming for an order of
magnitude estimate.

In Table \ref{tab:exp-data} we give, for a few materials, values of $C$, $t_d$, $f^+$ and $q_d$ extracted from experiments.

\renewcommand*{\arraystretch}{1.4}
\begin{table}[ht]
	\centering
	\begin{tabular}{@{}l  l  l  l  l@{}}
		\toprule
		& EuS & EuO & Fe & Ni \\ 
		\midrule
		$C$, $10^{- 3}$ & 16 & 5.2 & 0.13 & 0.040 \\ 
		$t_d$ & 0.31 & 0.14 & 0.010 & 0.0044 \\ 
		$f^+,\,${\r A} & 1.8 & 1.6 & 0.91 & 1.27 \\
		$q_d \hspace{0.17em}$, {\r A}$^{- 1}$ & 0.24 & 0.16 & 0.045 & 0.018\\
		\bottomrule
	\end{tabular}
	\caption{\label{tab:exp-data}We use $t_d \approx (4 \pi C)^{1 / 1.4}$ and
		$q_d = (4 \pi C)^{1 / 2} / f^+$. We give a detailed account of sources for
		this data in Appendix \ref{app:experiment}.}
\end{table}

\begin{figure}[ht]
	\centering
	\includegraphics[scale=1]{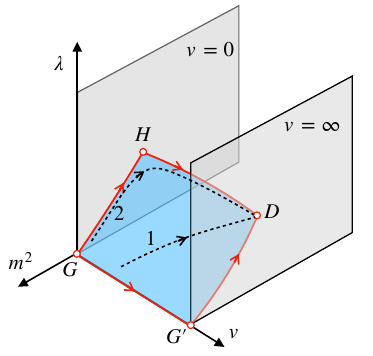}
	\caption{\label{fig-crossover}The RG flow diagram with four fixed points: $G$ - Gaussian, $G'$ - Gaussian dipolar, $H$ - Heisenberg, $D$ - dipolar. Materials with $t_d\sim 1$, like EuS and EuO, correspond to trajectories like 1 which flow straight to $D$. Materials with $t_d\ll 1$, like Fe and Ni, are supposed to correspond to trajectories like 2 which first approach $H$ and then flow to $D$. The shown trajectories correspond to the exact critical temperature $t=0$ (critical flow). For $t$ slightly different from 0, the RG flow initially tracks the critical trajectory, and then deviates from it. Depending on the value of $t$, this deviation may happen, for type 2 trajectories, when the flow is near $H$ or near $D$. This implies the crossover behavior mentioned in the main text.}
\end{figure}

We extract two conclusions from this table. First, the values of $t_d$ are much
larger for EuS and EuO than for Fe and Ni. Thus, we expect that EuS and EuO
will show only dipolar behavior, for $t \ll t_d$. On the other hand, as $t$ is lowered, Fe and Ni are expected to show Heisenberg behavior for $t_d \ll t \ll
1$, followed by a crossover to dipolar behavior for $t \ll t_d$. See Fig.~\ref{fig-crossover}.

Second, $q_d$ is also much smaller for Fe and Ni than for EuS and EuO. Thus,
much softer polarized neutrons will have to be used to see the suppression of
longitudinal fluctuations of the order parameter.

\subsection{Microscopic derivations of effective theory}

For Eu compounds, which are ferromagnetic insulators with well-localized magnetic moments, one can also give an independent estimate for parameters $a$ and $b$ starting from the microscopic Heisenberg model and performing Hubbard-Stratonovich transformation. The inputs in this computation are the critical temperature, the magnitude of the individual magnetic moments, and the lattice constant. This gives, for EuS and EuO, $a$ and $b$ in reasonable agreement with the estimates in Table \ref{tab:exp-data} extracted from the measurements of $\xi$ and $\chi$ (see App.~\ref{app:micro}). 

Early literature \cite{FisherAharony} attempted to estimate the strength of dipolar effects in ferromagnetic metals such Fe and Ni using similar microscopic arguments. Those estimates come out very different from Table \ref{tab:exp-data}, and we believe they cannot be trusted (reference \cite{PhysRevLett.51.833} also finds estimates for dipolar strength of Fe and Ni in tension with \cite{FisherAharony}). Indeed, the Heisenberg model description is not correct for Fe and Ni at the microscopic level, since their magnetic moments are not localized but are carried by electrons in the conductance bands (itinerant magnetism \cite{Kubler}). For such materials, direct microscopic estimates of $a$ and $b$ are bound to be much harder than for ferromagnetic insulators. On the other hand, the method leading to Table \ref{tab:exp-data} should be universally applicable.

The microscopic derivation given in App.~\ref{app:micro}, while not directly applicable to itinerant magnets, does illustrate an important point---that the unit normalization of the $-\phi\cdot \vec{B}^{(0)}$ coupling is inevitably related to the fixed normalization of the dipolar interaction term given in \eqref{Hpheno}. One could still ask, for the sake of the argument: what if we change the coefficient of the $\phi.U.\phi$ term from $1/2$ to $e/2$, with $e$ a new parameter? One change is that \eqref{eq:demag} would then become (see also \eqref{eq:phiHt})
\begin{equation}
	\phi = \frac{B^{(0)}}{b + 4 \pi D^{\rm eff}} = b^{-1} H^t,\qquad D^{\rm eff}= e D, \qquad H^t=B^{(0)}-4\pi D^{\rm eff}\phi\,.
\end{equation}
When measuring susceptibility, the value of $D^{\rm eff}$ is important when fitting the data. Thus any deviation of $D^{\rm eff}$ from the purely geometrically determined $D$ could be ascribed to $e\ne 1$. To our knowledge, no significant deviation is observed. E.g.~the works \cite{kotzler1986change,NoakesEtAl66} used the geometric $D=4\pi/3$ for spherical samples. Ref.~\cite{SeegerEtAl95} reports $N=0.32$ ($D=4\pi N$) for their spherical sample (sample No.1).

\subsection{Experimental evidence of the dipolar fixed point behavior}

The most dramatic effect of the dipolar fixed point is the suppression of longitudinal fluctuations of the order parameter. This effect can be seen by scattering polarized neutrons of the critical sample. For the longitudinal polarization, and for $t\ll t_d$, $q\ll q_d$, scattering cross-section should be suppressed. This effect was observed in Ref.~\cite{Kotzler1986} for EuS and EuO. For one of these materials (EuS) they also explored $q\sim q_d$ and saw that the longitudinal suppression disappears, in accord with the theory.

We next discuss measurements of the critical exponent $\gamma$ governing the asymptotic scaling behavior of the susceptibility $\chi\propto t^{-\gamma}$. In the Heisenberg fixed point, we have (using $\gamma=\nu(2-\eta)$ and the latest Monte Carlo measurements of $\nu$, $\eta$ from \cite{Hasenbusch-O3})\footnote{The latest conformal bootstrap result \cite{Chester:2020iyt} is in agreement but less precise $\gamma=1.3964(9)$.}
\begin{equation}
	\gamma_{H} = 1.39635(20)\,. \label{eq:gammaHeis}
\end{equation}
On the other hand for the dipolar fixed point the three-loop calculations of \cite{Kudlis:2022rmt} predict a smaller but very close value:
\begin{equation}
	\gamma_{D}= 1.381(8)\,.
\end{equation}
Ref.~\cite{PhysRevB.14.4908} measured $\gamma=1.387(36)$ for EuO and $\gamma=1.399(40)$ for EuS, in agreement with the theoretical value of $\gamma_D$.

It is interesting to consider the effective susceptibility exponent:
\begin{equation}
	\gamma_{\rm eff}=-\frac{d \log \chi}{d \log t}\,.
\end{equation}
For materials with $t_d\ll 1$, which show a crossover behavior from Heisenberg to dipolar, the effective exponent should be close to $\gamma_H$ for $t_d\ll t\ll 1$ and close to $\gamma_D$ for $t\ll d_d$, but it can deviate from these at $t\sim t_d$. The functional dependence of this deviation on $t/t_d$ is universal as it is controlled by the RG trajectory connecting the $H$ and $D$ fixed points (Fig.~\ref{fig-crossover}). This was computed to second order in the $\epsilon$-expansion by Bruce, Kosterlitz and Nelson \cite{Bruce_1976,Bruce_1977}, with a result that $\gamma_{\rm eff}$ should show a pronounced dip at $t\sim t_d$. This dip can be interpreted as follows. Let us write
\begin{equation}
	\chi\approx X_H t^{-\gamma_H}\qquad(t_d\ll t\ll 1),\qquad \chi\approx X_D t^{-\gamma_D}\qquad(t\ll t_d).
\end{equation}
Although we have $\gamma_H\approx \gamma_D$, the prefactors $X_H$ and $X_D$ do not have to be equal. The ratio $X_D/X_H$ is universal. If $X_D/X_H<1$, $\gamma_{\rm eff}$ will show a dip. This effect was confirmed experimentally by measurements on amorphous ferromagnets \cite{SrinathPRB,Srinath_2000}, reviewed in \cite{Kaul}.

\begin{remark} We would like to comment on the relatively recent susceptibility measurements in Ni \cite{SeegerEtAl95} which cover $5 \times 10^{-4} <t< 1.5\times 10^{-2}$. The estimate of $t_d$ for Ni in Table \ref{tab:exp-data} (and also in \cite{PhysRevLett.51.833}) falls in the middle of this interval. However, the observations of \cite{SeegerEtAl95} are inconsistent with this. 
	Indeed, Ref.~\cite{SeegerEtAl95} sees no sign of the Heisenberg to dipolar crossover; in fact, their $\gamma_{\rm eff}$ decreases monotonically when increasing $t$. The authors of \cite{SeegerEtAl95} assume that their $t$ belong to the $H$ critical region, and attribute the variation of $\gamma$ to corrections to scaling near the $H$ fixed point. Their experimental $\gamma$, extrapolated to $t=0$, is $\gamma=1.340(10)$, significantly lower than the theoretical value \eqref{eq:gammaHeis}. To explain this discrepancy, they hypothesize a long-range exchange interaction modifying the universality class. The theoretical origin of this ad hoc interaction is unclear.
	
	Experimental $\gamma$ lower than theory is typical for measurements in isotropic itinerant magnets (see e.g.~\cite{Hohenemser1989}, table 5). It would be interesting to understand why.
\end{remark}

\section{Scale invariance without conformal invariance}
\label{sec:scale}

We will now argue that the dipolar fixed point is scale invariant but not conformally invariant. We will give two arguments, one based on the 2pt function of $\phi_i$ and another on the form of the stress tensor and the existence of the virial current (including an explanation for its non-renormalization due to a shift symmetry). These discussions assume the $\epsilon$-expansion in $d=4-\epsilon$ dimensions, but some of the arguments such as non-renormalization of the virial current will be non-perturbative.

In Section \ref{sec:general} we list several other interacting models having scale without conformal invariance and identify shift symmetry as a general feature protecting the virial current dimension of all such currently known models.

\subsection{Two-point function argument}
\label{sec:2ptarg}
The simplest way to observe that the theory is not conformally invariant is to look at the 2pt function of $\phi_i$. This type of argument goes back to \cite{Dorigoni:2009ra,El-Showk:2011xbs}, and was also used in \cite{Mauri:2021ili}.

The 2pt function of $\phi_i$ with scaling dimension $\Delta_\phi$ is given by
\begin{align}
	\langle \phi_i(x) \phi_j(0) \rangle = \frac{A}{|x|^{2\Delta_\phi}} \left(\delta_{ij} - \alpha \frac{x_i x_j}{x^2} \right) \ ,\qquad \alpha = \frac{2\Delta_\phi}{2\Delta_\phi-(d-1)}\,, \label{2pt-phi}
\end{align}
where $\alpha$ is fixed by the transversality condition $\partial_i \phi_i = 0$. Unless $\Delta_\phi=d-1$,
this is different from the 2pt function of a primary vector field, which has $\alpha=2$. In $\epsilon$-expansion $\Delta_\phi=d-1$ can be excluded; indeed we have \cite{AF-PRBII}
\begin{equation}
	\Delta_\phi=\frac{d-2}{2}+\gamma_\phi,\qquad \gamma_\phi =\frac{10}{867} \epsilon^2+O(\epsilon^3)\,,
	\end{equation}
and the anomalous dimension $\gamma_\phi$ remains tiny in any $3\le d\le 4$; the recent $d=3$ calculation \cite{Kudlis:2022rmt} found $\eta_{\rm dip}=2\gamma_\phi= 0.033(8)$. Hence $\phi_i$ cannot be a conformal primary. Being the lowest dimension field of the theory, it cannot be a descendant either, finishing the proof. 

As mentioned in the introduction, often questions of scale and conformal invariance are studied imposing unitarity \cite{Polchinski:1987dy,Dorigoni:2009ra,Luty:2012ww,Dymarsky:2013pqa,Dymarsky:2014zja}. In this respect, it is worth pointing out that the dipolar fixed point is not unitary.
 Consider the 2pt function \eqref{2pt-phi} with the separation being in the $x_1$ direction, i.e. $x=(1,0,0,\ldots)$. In this configuration
 we have, from \eqref{2pt-phi}:
 \begin{align}
 	 &\langle \phi_1(x) \phi_1(0) \rangle =A(1-\alpha)\,,\\
  &\langle \phi_i(x) \phi_i(0) \rangle =A \qquad (i\ne 1)\,,  
 \end{align}
 while reflection positivity demands that
 \begin{align}
 	&\langle \phi_1(x) \phi_1(0) \rangle \le 0\,,\\
 	&\langle \phi_i(x) \phi_i(0) \rangle \ge 0 \qquad (i\ne 1)\,.
 	\end{align}
 We see that these constraints require $\alpha \ge 1$, that is $\Delta_\phi \ge  \frac{d-1}{2}$, or equivalently $\gamma_\phi \ge 1/2$, which is excluded by the above perturbative estimates of this anomalous dimension.\footnote{This argument also shows that the transverse vector unparticle proposed originally by Georgi in \cite{Georgi:2007ek} is ruled out  in the range $\Delta_\phi < \frac{3}{2}$ even if we assume the existence of scale-invariant but non-conformal field theories in $d=4$.} 
 
 \begin{remark}
As we have seen in Section \ref{sec:locdesc}, the dipolar model Hamiltonian \eqref{effLagr} can be obtained by coupling the $O(3)$ model to the fluctuating magnetic field, \eqref{eq:HB}, integrating out the magnetic field and taking the low energy limit. The $O(3)$ model and the magnetic field Hamiltonian are separately unitary. So how can their coupling produce a non-unitary theory?\footnote{We thank Juan Maldacena for raising this question and answering it.} The answer is that unitarity is broken by the coupling term, $\phi_i B_i$ in \eqref{eq:HB}, which treats $\phi_i$ as a vector field, while it was a scalar multiplet in the $O(3)$ model.
 	\end{remark}
 	
\begin{remark}
	An equivalent way to study the unitarity of the 2pt function \eqref{2pt-phi} is by considering the Wightman spectral density in momentum space \cite{Grinstein:2008qk}, which can be obtained as the imaginary part of the Schwinger function in momentum space \eqref{2ptmom}, continued from Euclidean to Lorentzian. We have:
 \begin{equation}
 	\langle \phi_i(q) \phi_j(-q)\rangle \propto \theta(q^0) \theta(q^2) \frac{1}{(q^2)^{2-\gamma_\phi}} \left[q^2 \eta_{i j} - q_i q_j\right]\,,
 	\label{2ptW}
 \end{equation}
 where we use Lorentzian notation with the mostly minus metric. The Wightman function is supported in the forward cone $q^0\ge 0$, $q^2\ge 0$. 
  
  Multiplying \eqref{2ptW} by external wavefunctions $\chi_i(q)$ and $\chi_j^*(-q)$ and integrating over $q$, unitarity requires that the answer should be nonnegative. The expression in brackets gives
  \begin{equation}
  	q^2 \chi.\chi^* - |\chi.q|^2 \,,
  	\end{equation}
  which is non-negative inside the forward cone, as can be seen going to the rest frame $\vec q=0$, where it becomes $q_0^2 |\vec \chi|^2$. The lack of unitarity of this theory comes not from the negativity of the Wightman function inside the forward cone (it is positive), but from the behavior of the integrand near the null cone. Indeed a positive distribution must be a \emph{measure}, i.e.~integrable. The strongest constraint comes from approaching the cone transversally and imposes the constraint (cf \cite{Grinstein:2008qk}, Eq.~(4.7)):
  \begin{equation}
  	\text{unitarity} \Longrightarrow \gamma_\phi\ge 1,
  	\end{equation}
which is even stronger than the condition $\gamma_\phi\ge 1/2$ found above.\footnote{{Without extra assumptions, one cannot improve the bound further due to the existence of a concrete example: take a free massless scalar $\varphi$ and consider $\phi_i=\partial_i \varphi$ (so that $\gamma = 1$ in \eqref{2ptW}). It is conserved because of the equation of motion, and the 2pt function must be consistent with the unitarity bound.}} However, the $x$-space argument was providing only the necessary condition, since we only examined reflection positivity for the field $\phi_i$. Considering derivatives of $\phi_i$, it should be possible to improve the $x$-space argument and rule out the range $1/2\le \gamma_\phi <1$.
 \end{remark}

\begin{remark}
	Note that the the position space correlator \eqref{2pt-phi} becomes singular ($\alpha=\infty$) for $\Delta_\phi = \frac{d-1}{2}$.
One wonders if this could lead to a nonperturbative argument that $\Delta_\phi$ cannot cross $\frac{d-1}{2}$ between UV and IR. However, there does not seem to be a simple way to show this. Using an ansatz allowing for violations of scale invariance at intermediate scales:
	 \begin{align}
	 	\langle \phi_i(x) \phi_j(0) \rangle = f(x^2) \left(\delta_{ij} + g(x^2) \frac{x_i x_j}{x^2} \right) \,, \label{2pt-phi-NP}
	 \end{align}
 the constraint $\partial_i \phi_i =0$ imposes:
 \begin{equation}
 	\frac{dg}{d\rho}  = \frac{d\log f} {d\rho} +\left[ \frac {1-d}2 - \frac{d\log f} {d\rho}\right]g\,,\quad \rho=x^2\,.
 	\end{equation}
 If $\frac{d\log f} {d\rho}=(1-d)/2$ for some $\rho=\rho_0$, the second term in the r.h.s.~vanishes. This implies that $\frac{dg}{d\rho} $ must be nonzero at such $\rho$, but does not signal any particular singularity in integrating the equation. 
	\end{remark}

  \subsection{Stress tensor argument}
  \label{sec:stress-tens-arg}
  The second, classic \cite{Polchinski:1987dy}, way to understand whether a fixed point is scale invariant with or without conformal invariance, proceeds via the properties of the trace of the stress tensor, which is a response to the background metric.
  According to Polchinski's analysis \cite{Polchinski:1987dy}, a local fixed point is scale invariant if the trace of the stress tensor $T_{ij}$\footnote{We always assume that the stress tensor is symmetric to make the rotational invariance manifest.} is given by the divergence of a vector operator $V_i$, referred to as the virial current:
  \begin{align}
    \label{Tscale}
    T_{ii} = -\partial_i V_i \, .
  \end{align}
Furthermore, the fixed point is conformally invariant (in $d>2$, which is our case of interest here) if in addition to \eqref{Tscale}, the virial current is given by a divergence of a local operator, namely
  \begin{equation}
    \label{Tconf}
    V_i = \partial_j \mathcal{O}_{ij} \, ,
    \end{equation}
where $\mathcal{O}_{ij}$ can be assumed symmetric without loss of generality. We will call virial currents satisfying this condition \textit{improvable}. If it holds, an ``improved'' stress tensor can be found which is traceless \cite{Polchinski:1987dy}.

  To discuss the stress tensor in our model, we start from the local effective Hamiltonian \eqref{effLagr}, which eliminates the longitudinal fluctuations of the order parameter via the Lagrange multiplier.
  We will see below that our fixed point has a virial current $V_i \propto U \phi_i+\ldots$ where $\ldots$ stands for improvable terms. 
  
  Here we present the gist of the argument, postponing to Appendix \ref{sec:renorm} a more detailed treatment of the effects of renormalization.
  To compute the stress tensor in $d = 4-\epsilon$ dimensions, it is convenient to redefine $U \to U + \frac12 \partial_i \phi_i$, which gives an effective action equivalent to \eqref{effLagr}:
  \begin{align}
  \label{effLagr2}
    \widetilde{\mathcal{H}}
    = \int d^d x \left(\frac{1}{4}f_{ij}^2
    + \phi_i \partial_i U
    + \frac{\lambda}{4}(\phi_i\phi_i)^2 \right) \, .
  \end{align}
Here $f_{ij} = \partial_i \phi_j - \partial_j \phi_i$, and we have already assumed that $m^2$ is at the critical value, which is $m^2=0$ in dimensional regularization. The EOM are
  \begin{align}
  	\label{eom2}
    \partial_i f_{ij} = \partial_j U + \lambda (\phi_i^2) \phi_j \ ,\qquad \partial_i \phi_i=0\,.
  \end{align}
  The (classical) stress tensor is computed as usual by varying the background metric. The merit of using $f_{ij}$ is that we have no covariant derivative in the action.\footnote{\label{note:covariant}We vary  $\int d^d x\sqrt{g} \left( \frac{1}{4} g^{ij} g^{kl} f_{ik}f_{jl} + g^{ij} \partial_i U \phi_j + \frac{\lambda}{4} (g^{ij} \phi_i \phi_j)^2 \right)$. Stress tensors computed from \eqref{effLagr} and \eqref{effLagr2} differ by improvable and EOM terms.} We only give here the expression for the trace of the stress tensor (see App.~\ref{sec:renorm} for the full stress tensor):
  \begin{align}
   T_{ii}
   &= \frac{\epsilon}{4}\big( f_{ij}^2 + \lambda (\phi_i^2)^2 \big)
    + (2-d) \phi_i \partial_i U \notag \\
   &= - \frac{\epsilon}{4} \lambda (\phi_i^2)^2
    - \frac{d}{2} \partial_i \left(  \phi_i U \right) + \frac{\epsilon}{2} \partial_i \partial_j \left( \frac12 \phi^2 \delta_{ij} - \phi_i \phi_j \right) \, 
    \label{eq:Tii}.
  \end{align}
  In the second line, we used the EOM to express the trace as a sum of a quartic term, a virial current and an improvable term.
  To include the effects of renormalization, we should express the trace in terms of renormalized fields.
  We carry this out in Appendix \ref{sec:renorm}, and here just give the main results.
  For the quartic term, renormalization amounts to the replacement $\epsilon \lambda (\phi_i^2)^2 \to - \beta(\lambda) (\phi_i^2)^2 -4\gamma_\phi \partial_i \left(  \phi_i U \right) + \ldots$, with $\ldots$ improvable terms. At the fixed point, the beta-function vanishes $\beta(\lambda)=0$ and the quartic term drops out.
We finally obtain 
  \begin{equation}
  	\label{eq:V0}
    T_{ii} \, \big|_{\text{fixed point}}
    = -\partial_i V_i, \qquad V_i
    = V^{(0)}_i  + \partial_j \mathcal{O}_{ij}\,,\qquad V^{(0)}_i  = \Delta_U \, U\phi_i \,.
    \end{equation}
    The coefficient $\Delta_U$ in the last equation is shifted from the classical value $d/2$ to the renormalized value $\Delta_U=d/2-\gamma_\phi$.\footnote{The formal derivation of this finite shift can be found in Appendix \ref{sec:renorm}. Here we would like to point out the following amusing connection. We know that $\Delta_\phi=d-1-\Delta_U$. If $\Delta_U$ were zero (which does not happen in the dipolar model, because $\gamma_\phi$ is tiny), $\Delta_\phi$ would become consistent with the conformal symmetry, as we have seen in Section \ref{sec:2ptarg}. We see that this goes hand in hand with the vanishing of the unimprovable part $V^{(0)}$ of the virial current in \eqref{eq:V0}.}
The part $V^{(0)}_i$ of the virial current is not conserved: using EOM \eqref{eom2}, we have $\partial_i V^{(0)}_i = \Delta_U  \, \partial_i U \phi_i \ne 0$. It is also not improvable, as it is built of fields that carry no derivatives, while according to \eqref{Tconf}, improvable virial currents should contain at least one derivative.

  Thus we arrive at the same conclusion as in Section \ref{sec:2ptarg}: the dipolar fixed point is an interacting scale invariant fixed point without conformal invariance.

\subsection{Virial current dimension and the shift symmetry}
\label{sec:shift}

This result begs the following question. Eq.~\eqref{Tscale} means that the
virial current operator $V_i$ has scaling dimension exactly $d - 1$, since the
stress tensor has scaling dimension $d$.\footnote{To be precise, an
	arbitrarily chosen stress tensor may not have a well-defined scaling
	dimension, but it was shown in {\cite{Polchinski:1987dy}} how to find an
	``improved'' stress tensor having canonical scaling dimension $d$ (see also
	{\cite{Nakayama:2013is}}). In this section, we use such a stress tensor and the
	corresponding virial current which has scaling dimension $d - 1$. This scaling
	virial current may differ from the virial current in the previous section by
	improvable terms. We review these arguments in Appendix \ref{sec:scaling}.} In other words, its scaling dimension is not renormalized
from the canonical dimension $d - 1$. Usually, the only vector operators that
are not renormalized are the conserved currents, making scale without
conformal invariance generically impossible in the presence of interactions
{\cite{Rychkov:2016iqz,El-Showk:2011xbs}}. Yet $V_i$ is definitely not
conserved. There must be something non-generic about the dipolar fixed point,
allowing $V_i$ to not renormalize in the presence of interactions.

This non-generic feature is the \textit{shift symmetry} of model
\eqref{effLagr}. The shift symmetry acts on the fundamental field $U$ by $U
(x) \to U (x) + u$ where $u$ is a constant. The origin of this symmetry is
the Bianchi identity of the $B$ field.

The shift symmetry is a global symmetry group, $G_{\text{shift}}
=\mathbb{R}$. In particular, it commutes with Poincar{\'e}. We can also
consider the shift symmetry charge $Q$, which by definition acts on the
fundamental field as
\begin{equation}
	[Q, U (x)] = 1\qquad(\text{i.e. } \delta_Q U=1) \, .
\end{equation}
Therefore the scaling dimension of $Q$ is given by
\begin{equation}
	\Delta_Q = - \Delta_U \, . \label{QU}
\end{equation}
The conserved current generating the shift symmetry is $\phi_i$. The charge
$Q$ can be obtained as a surface integral of the conserved current, $Q = \int
d \Sigma_i  \hspace{0.17em} \phi_i$, from where we get an alternative
expression for its scale dimension:
\begin{equation}
	\Delta_Q = \Delta_{\phi} - (d - 1) \, .
\end{equation}
The two expressions for $\Delta_Q$ are consistent by \eqref{phiU}. This also
provides an alternative way of understanding \eqref{phiU}.

Eq.~\eqref{QU} can also be equivalently written as a commutation relation
between $Q$ and the dilatation generator $D$:
\begin{equation}
	[D, Q] = - \Delta_U Q \,. \label{DQ}
\end{equation}
Since $\Delta_U > 0$ in our model, this equation means that $Q$ acts as a
lowering operator for the scaling dimension. Scaling operators in the dipolar
fixed point will come in infinite ``shift multiplets'':
\begin{equation}
	\label{ShiftMult}
	 \{ \mathcal{O}_0, \mathcal{O}_1, \ldots \} \, ,
	 \end{equation}
where
\begin{eqnarray}
	[Q, \mathcal{O}_n] & = & \mathcal{O}_{n - 1} \quad (n \geqslant 1) \, , \qquad
	[Q, \mathcal{O}_0] = 0  \label{multipl}
\end{eqnarray}
and
\begin{equation}
	\label{deltaOn}
		\Delta  (\mathcal{O}_n)  = \Delta  (\mathcal{O}_0) + n \Delta_U .
\end{equation}
Generally, in free theory we have $\mathcal{O}_n = \frac{1}{n!} U^n\mathcal{O}_0$. However, our construction of shift multiplets does not rely on perturbation theory, and the scaling dimensions \eqref{deltaOn} are valid non-perturbatively.

Let us focus on the shift multiplet constructed on top of $\mathcal{O}_0 =
\phi_i$. We have:
\begin{equation}
	\label{deltaO1}
	\Delta (\mathcal{O}_1) = \Delta_{\phi} + \Delta_U = d - 1 .
\end{equation}
I.e. $\mathcal{O}_1$ has scaling dimension $d - 1$, which is what we need!
However, it's not a conserved current. Indeed, in free theory we
have
\begin{equation}
	(\mathcal{O}_1)_i = U \phi_i .
\end{equation}
At the interacting IR fixed point we will have
\begin{equation}
	\label{O1form}
	(\mathcal{O}_1)_i = U \phi_i + \ldots
\end{equation}
where $\ldots$ are terms of 4d scaling dimension 3 which have schematic
form $\partial \phi^2$. These terms need to be added to $U \phi_i$ to make it a
good scaling operator. Note that {\ldots} terms do not involve $U$ because the
mixing matrix of $U \phi_i$ and $\partial \phi^2$ is triangular: $\partial
\phi^2$, being neutral under the shift symmetry, cannot generate $U
\phi_i$ under RG flow.\footnote{The operator $\partial_i U$ also has 4d scaling dimension 3, but being $\mathbb{Z}_2$ odd it does not appear in $\ldots$.}

We have shown that the dipolar fixed point contains a vector operator $(\mathcal{O}_1)_i$ of the form \eqref{O1form}, having scaling dimension exactly $d-1$. This achieves the main goal of this section - to show how the shift symmetry may naturally provide non-conserved operators of this scaling dimension.

The virial current $V_i$ is also of the form \eqref{O1form}, up to $\Delta_U$ rescaling. Let us consider the stress tensor which has a well-defined scaling dimension ($d$), so that the corresponding virial current, which we denote $V^{\text{scale}}$ also has a well-defined scaling dimension ($d-1$). This also fixes the $\ldots$ terms in the virial current. It can be shown that the $\ldots$ terms in $V^{\text{scale}}$ and $\mathcal{O}_1$ are the same (up to $\Delta_U$ rescaling), that is:
\begin{equation}
V=\Delta_U	\mathcal{O}_1\,,
	\end{equation} 
This should not be surprising - it is already difficult to have \emph{one} non-conserved vector of scaling dimension exactly $d-1$; to have \emph{two} would be inexplicable. Thus we will not verify this explicitly here.

We will see in Section \ref{sec:general} that shift symmetries are at work not just for the dipolar fixed point but for all known interacting models of scale without conformal invariance. In all of them, the virial current dimension can be seen protected by a shift symmetry.

\begin{remark}
	For completeness, we give here the original version of the argument for the non-renormalization of the virial current \cite{ERG-talk}. There, the main idea was to exploit the shift symmetry by considering the 2pt functions $\langle V_j (x_2) \phi_k(x_3) \rangle$ and $\langle \phi_j(x_2) \phi_k(x_3) \rangle$, which are related by shift symmetry:
\begin{equation}
	\delta_Q \langle V_j (x_2) \phi_k(x_3) \rangle = \Delta_U\langle \phi_j(x_2) \phi_k(x_3) \rangle\,.
	\end{equation} 
This relation can be written as the following Ward-Takahashi identity involving the divergence of $\phi_i$, which is the shift symmetry current: 
\begin{align}
	\langle \partial_i \phi_i(x_1) V_j (x_2) \phi_k(x_3) \rangle = \Delta_U \delta^{d}(x_1-x_2) \langle \phi_j(x_2) \phi_k(x_3) \rangle  \ .
\end{align}
Equating the scaling dimensions of both sides gives
\begin{equation}
	\Delta_\phi+1+\Delta_V+\Delta_\phi = d+ 2\Delta_\phi\ \Longrightarrow\ \Delta_V=d-1\,.
	\end{equation}
This argument for the nonrenormalization of $\Delta_V$ is basically equivalent to the argument given in the main text. Although it is shorter, at the first look it may appear a bit ad hoc. The argument in the main text, based on equations \eqref{DQ}, \eqref{deltaO1}, is hopefully useful to understand the inner workings of this mechanism.
\end{remark}

\begin{remark} It is amusing to recall that the usual argument for the non-renormalization of the conserved current dimension is also based on a Ward-Takahashi identity. Namely, we have, for a conserved current $J$ associated with a linearly realized global symmetry, a Ward-Takahashi identity of the schematic form $\langle \partial_i J_i(x_1) \varphi (x_2) \ldots) \rangle \propto \delta^{d}(x_1-x_2) \langle \varphi (x_2) \ldots \rangle$, which implies $\Delta_J=d-1$.
	\end{remark}

\subsection{Other consequences of shift symmetry}\label{OC}

Let us discuss an additional role of the shift symmetry concerning the $U^2$ operator. This operator is classically marginal. Taking into account interactions, it will become weakly relevant (see below), and could destabilize the dipolar fixed point. However, since the operator is charged under the shift symmetry, it is not generated by the RG, and so the fixed point is protected.

Let us discuss the dimension of $U^2$ in more detail. Let us call by $[U^2]$ the renormalized $U^2$ operator, which differs from $U^2$ by pieces of the schematic form $(\partial \phi)^2$  and $\partial (\phi U)$, with which it can mix under renormalization. The operator $[U^2]$ can be equivalently defined as the first excited member $\mathcal{O}_1$ of the shift multiplet \eqref{ShiftMult} built on top of $\mathcal{O}_0=U$. Therefore, we have
\begin{equation}
	\Delta_{[U^2]} = 2\Delta_U = 2(d-1-\Delta_{\phi})=d -2\gamma_\phi\,,
	\end{equation}
where the first equation follows from \eqref{deltaOn}. Since $\gamma_\phi$ is positive, the $U^2$ deformation is indeed relevant, as stated above.

As we said, $U^2$ is not going to be generated by the RG starting from \eqref{effLagr}. But what will happen if we add it by hand, thus breaking the shift symmetry?
It is reasonable to guess that this deformation will start a flow which, if the mass term is properly perturbed as well, will eventually take us back to the $O(d)$ Wilson-Fisher fixed point, where the theory is conformally invariant. We can test this guess for consistency by looking at what happens at the end point of this flow, when the Wilson-Fisher fixed point is approached. Around the $O(d)$ Wilson-Fisher fixed point (or $O(3)$ Heisenberg fixed point in three dimensions), the flow will be induced by the leading deformation which breaks the $O(d)$ spatial times $O(d)$ internal symmetry of Wilson-Fisher to the diagonal $O(d)$. This deformation $\mathcal{O}$ can be constructed from the Wilson-Fisher primary $R_{\mu\nu,ij}$ of the schematic form $\phi_i \partial_\mu \partial_\nu \phi_j$, which is spin two in both space-time and $O(d)$, by contracting indices appropriately: $\mathcal{O}= \delta_{\mu i}\delta_{\nu j} R_{\mu\nu,ij}$. It is known that in $d=4-\epsilon$ \cite{Wilson:1973jj}
\begin{equation}
	\Delta_\mathcal{O}=d+\frac{d}{(3(d+8))^2}\epsilon^2\,.
	\end{equation} 
This anomalous dimension has to be positive because this primary is not conserved, and unitarity implies that a non-conserved spin-two operator has dimension strictly greater than $d$. So $\mathcal{O}$ is indeed irrelevant, as it should be if the flow leads from the dipolar to the Wilson-Fisher fixed point.

We see that once we break the shift symmetry, we flow to a fixed point which is conformal.
This may be traced back to the need to have the virial current of dimension $d-1$. Once the shift symmetry is broken, there is nothing that protects the dimension of the virial current in an interacting theory. We will see in the next section that all known scale without conformal fixed points have a shift symmetry.

\begin{remark}
At the beginning of the paper, we discussed a flow that takes us from Wilson-Fisher to dipolar via a non-local deformation \eqref{eq:Vdip}. The non-local deformation could be obtained, as in Section \ref{sec:locdesc}, by coupling Wilson-Fisher to another local sector (magnetic field). This additional local sector is not reproduced when we get back from dipolar to Wilson-Fisher via a local deformation breaking the shift symmetry, as discussed above. Thus the RG flow is not circular.
	\end{remark}

\section{Other interacting models with scale without conformal invariance}
\label{sec:general}

In this section, we discuss several other interacting scale without conformal models.
An important feature of all these models is that they have a shift symmetry that acts on some fundamental field as $\Om(x) \to \Om(x) + c$, with $c$ a constant.
It is this symmetry that prevents scale invariance from getting enhanced to conformal invariance.

To see that shift-invariant models are not conformal one simply looks at the 2pt function of the shift current.
This is in fact the strategy of Section \ref{sec:2ptarg}, because $\phi_i$ is the shift current in the dipolar model.
Throughout this section we call $J_\mu$ the shift current, and we call $Q = \int d\Sigma_\mu J_\mu$ its charge.
In this notation, conservation of the shift current fixes its 2pt function
\begin{align}
 \langle J_\mu(x) J_\nu(0) \rangle
 = \frac{A}{|x|^{2\Delta_J}} \left(\delta_{\mu\nu} - \frac{2\Delta_J}{2\Delta_J-(d-1)} \frac{x_\mu x_\nu}{x^2} \right) \, .
 \label{2pt-JJ}
\end{align}
If $J_\mu$ is a primary field, that is if $J_\mu$ is not a total derivative of another local operator, then this is compatible with conformal invariance only when $\Delta_J = d-1$.
However, the way shift symmetry acts on $\Om$ implies
\begin{align}
[Q, \Om(x)] = \delta_Q \Om(x) = 1
\qquad \Longrightarrow \qquad
\Delta_J = d-1-\Delta_\Om  \, .
\label{eq:dimcurr}
\end{align}
As a result, if $J_\mu$ is a primary field and $\Delta_\Om\ne0$, the fixed point is scale but not conformally invariant.\footnote{The only subtlety is that the shift current could be a descendant. Since $\delta_Q\Om = 1$ we get the 2pt function $\langle J_\mu(q) \Om(-q)\rangle = iq_\mu/q^2$, and conformal invariance would require $J_\mu$ to be a descendant of $\Om$. In this case $J \sim \partial^n \Om$ and $\Delta_J = \Delta_\Om + n$.
Compatibility with \eqref{eq:dimcurr} requires $\Delta_\Om = \frac{d-1-n}{2}$ for some integer $n$.
This scenario is realized in free theories with Lagrangian $\Lm = \frac{1}{2} (\partial_{\mu_1} \ldots \partial_{\mu_m} \Om)^2$.
However, in interacting theories generically $\Delta_\Om \ne \frac{d-1-n}{2}$ and only scale invariance will be realized.}
This elementary argument explains why all models below, and more generally any interacting shift-invariant theory, are scale but not conformally invariant.

This still begs the question of why all these models have a virial current $V_\mu$ with dimension exactly $\Delta_V = d-1$.
The main idea is the same as in Section \ref{sec:shift}, namely that shift symmetry provides a candidate virial current $\Vm_\mu$ with the right dimension.
The property that defines $\Vm_\mu$ is that under shift-symmetry it maps to the current $J_\mu$, namely
\begin{equation}
 \label{eq:Vcand}
 [Q, \mathcal{V}_\mu] = \delta_Q \mathcal{V}_\mu = J_\mu
 \quad \Longrightarrow \quad
 \Delta_\mathcal{V}=d-1\,.
\end{equation}
The implication $\Delta_{\mathcal{V}}=d-1$ is valid assuming that $\Vm_\mu$ is a good scaling operator (i.e.~that it has a well-defined scaling dimension). We stress that $\mathcal{V}_\mu$ is a \emph{candidate} virial current. To guarantee that only scale invariance is present, it is still necessary to check that this $\mathcal{V}_\mu$ is the true virial current $V_\mu$, i.e.~that it does appear in the trace of the stress tensor, and moreover that it is not improvable.
The goal of the rest of this section is to show in detail how this mechanism works for several models of interest.
The argument in \eqref{2pt-JJ}-\eqref{eq:dimcurr} trivially applies to all these models, and shall not be repeated below.

\subsection{Landau-gauge massless QED in \texorpdfstring{$d=4-\epsilon$}{d=4-eps}}

	We start by reviewing the examples of \cite{Nakayama:2016cyh}. Historically, these were the first interacting examples of scale without conformal invariance. The first example is the Landau-gauge massless QED in $d=4-\epsilon$ dimensions. It is known that massless QED (a $U(1)$ gauge field + a massless fermion) flows to a fixed point in $d=4-\epsilon$ dimensions, which is conformal in the gauge-invariant sector \cite{Giombi:2015haa,Chester:2016ref}. Let us consider the gauge-fixed action for the same flow, in the Landau gauge, implemented via a Lagrange multiplier:
\begin{align}
S = \int d^d x \left(-\frac{1}{4 e^2}(\partial_\mu A_\nu-\partial_\mu A_\nu)^2 + B \partial_\mu A_\mu + i\bar{\psi} D_\mu \gamma_\mu \psi
 + \text{ghosts}  \right) \ .
 \label{eq:qed}
\end{align}
Usually, the gauge fixed action is treated as a formal device to do computations for the original theory. Here, following \cite{Nakayama:2016cyh}, we consider it is as a field theory in its own right. It is a well-defined field theory, albeit non-unitary.
The decoupled ghost sector is necessary for BRST invariance.

The bosonic part of the action \eqref{effLagr2} bears some similarity with the dipolar model action \eqref{effLagr2}, except for the absence of the quartic term $(A_\mu A_\mu)^2$. Of course, this term was absent in the gauge theory from which \eqref{eq:qed} originated. At the level of theory \eqref{eq:qed}, this term is forbidden by the BRST invariance of \eqref{eq:qed}. In the dipolar model \eqref{effLagr2}, there was no BRST invariance, and the quartic term was allowed.

Now, the main points of \cite{Nakayama:2016cyh} are:
\begin{itemize}
	\item
Theory \eqref{eq:qed} has a fixed point at the same value of the gauge coupling as the original gauge-invariant theory.
\item
This fixed point is scale invariant but not conformal invariant, with the non-zero virial current $V_\mu = (d-2)B A_\mu$.
\item
There is no contradiction with the conformal invariance of the gauge theory fixed point. Indeed, the virial current together with the decoupled ghost contribution is BRST trivial, so after taking the BRST cohomology, the theory does become conformally invariant.
 \end{itemize}

We can now see that the dimension of the virial current in this model is protected by the same mechanism \eqref{eq:Vcand}. The shift symmetry $Q$ is $B \to B + b$ for $b$ constant; it is generated by the current $J_\mu = A_\mu$. We have $\delta_Q V_\mu \propto J_\mu$. Thus $\Delta_ V=d-1$, as pertains to the virial current.

In \cite{Nakayama:2016cyh}, the non-renormalization of the virial current was proved in a slightly different manner by using the BRST symmetry. Our derivation here is more direct because we do not refer to the decoupled ghost sector.

\subsection{Landau-gauge Banks-Zaks fixed point in \texorpdfstring{$d=4$}{d=4}}

The next example from \cite{Nakayama:2016cyh} applies the same idea to non-abelian gauge theories in 4d (as opposed to $d=4-\epsilon$). In 4d, we know infinitely many examples of gauge theories with massless matter that show non-trivial fixed points such as the Banks-Zaks fixed points or the $\mathcal{N}=4$ super Yang-Mills theory. These fixed points show conformal invariance in the gauge-invariant sectors, but the corresponding gauge-fixed theories show scale invariance without conformal invariance \cite{Nakayama:2016cyh}. However, after taking the BRST cohomology and thus restricting to the gauge-invariant sector, these scale invariance fixed points become conformal.

In more detail, consider the Yang-Mills theory with massless matter. Here as in the previous section we restrict to the Landau gauge (see the next section for an example not in the Landau gauge). The action is:
\begin{align}
	\label{eq:YM}
S = \int d^4x \left( -\frac{1}{4 g^2} F_{\mu\nu}^a F_{\mu\nu}^a + B^a \partial_\mu A_\mu^a + i\bar{c}^a \partial_\mu D_\mu c^a + \mathrm{matter} \right) \ .
\end{align}

Let us assume that the gauge coupling, as well as the other matter coupling constants if any, flow to a fixed point. In this general setup, the analysis of the stress tensor trace shows \cite{Nakayama:2016cyh} that theory \eqref{eq:YM} is scale invariant but not conformal. The virial current is
\begin{equation}
	\label{eq:virialYM}
	V_\mu = \Delta_B (B^a A_\mu^a + i \bar{c}^a D_\mu c^a) \,.
	\end{equation}
Here the coefficient in the virial current is shifted from the classical value of $2$ to  $\Delta_B = 2 -\gamma_A $.
While the full theory is only scale invariant, conformal invariance would be recovered were we to restrict to the BRST-invariant subsector. This is because the above virial current is BRST trivial:
\begin{equation}
	\label{eq:VBRST}
V_\mu = \{Q_{\mathrm{BRST}}, \Delta_B \bar{c}^a A^a_\mu \} \,.
\end{equation} 
Thus, within the BRST cohomology, the energy-momentum tensor becomes traceless. This is why scale invariance of the gauge-fixed theory \cite{Nakayama:2016cyh} is not in contradiction with conformal invariance of the corresponding gauge-invariant fixed point. 

Now let us discuss how the dimension of the virial current is protected.
In addition to the BRST invariance, theory \eqref{eq:YM} has \emph{two} shift symmetries $B^a \to B^a + \lambda^a$ and $\bar{c}^a \to \bar{c}^a + \bar{b}^a $ with constant $\lambda^a, \bar{b}^a$.\footnote{The second shift symmetry exists because the ghost $c^a$ is independent of the anti-ghost $\bar{c}^a$. Not all textbooks treat this point properly.} Although we won't need it, we note the algebra satisfied by their charges:
\begin{equation}
	\label{eq:QQQ}
	[Q_{\rm BRST}, Q_B]=Q_{\bar c},\qquad  \{Q_{\rm BRST}, Q_{\bar c}\}=0,\qquad [Q_B,Q_{\bar c}]=0\,.
	\end{equation}
The currents of the $B$ and $\bar c$ symmetries are the fields $A_\mu$ and $D_\mu c$. Moreover we have:
\begin{equation}
[Q_B, B A_\mu] = A_\mu,\quad [Q_{\bar c}, \bar c\, D_\mu c] = D_\mu c\,.
\end{equation}
Both these equations are of the form \eqref{eq:Vcand}. Applying the general argument, we conclude that theory \eqref{eq:YM} contains \emph{two} fields of dimension $d-1$, namely the level-1 fields of the $B$-shift and $\bar c$-shift symmetry multiplets built on top of the corresponding shift currents $A_\mu$ and $D_\mu c$. The virial current \eqref{eq:virialYM} is their particular linear combination, so it also has dimension $d-1$.

This example produces infinitely many interacting scale invariant but non-conformal field theories where the dimension of the virial current is protected by the shift symmetries. 

\subsection{Fixed points not in the Landau gauge}
The third example from \cite{Nakayama:2016cyh} is a generalization of \eqref{eq:YM} from the Landau gauge to a general $\xi$ gauge, i.e.~adding the term $-\frac{1}{2\xi} (B^a)^2$ to the action. The gauge-parameter $\xi$ is then treated as a dimensionless coupling that runs under the RG flow. Then the Landau-gauge value $\xi=\infty$ is always a fixed point. Depending on the theory, there may exist other fixed points of the gauge parameter $\xi$. In QED, there is no other fixed point than the Landau gauge, but in non-Abelian theories, such fixed points were found in \cite{Nakayama:2016cyh}.

The virial current is still given by Eqs.~\eqref{eq:virialYM}, \eqref{eq:VBRST}. We would like to explain that its dimension is $d-1$. Since $\xi\ne\infty$, we no longer have the shift symmetry of $B^a$, so the argument from the previous section does not apply. However, we can still give a robust argument, using the shift symmetry of $\bar{c}^a$, in combination with the BRST invariance.

The argument is based on the following four equations:\footnote{Unlike in \eqref{eq:VBRST}, the coefficient $\Delta_B=2-\gamma_A$ in \eqref{q1} remains at its classical value 2, because $\gamma_A=0$ at this fixed point. This is because the form of the beta-function equation for $\alpha=1/\xi$ is \cite{Caswell:1974gg} $\beta_\alpha=\alpha \gamma_A$. So the fixed points with $\alpha=0$, like in the previous subsection, may have $\gamma_A\ne0$, while the fixed points with $\alpha\ne 0$ should have $\gamma_A=0$ \cite{Nakayama:2016cyh}.} 
\begin{align}
	V_\mu &= [Q_{\rm BRST}, \Delta_B\, \bar c\, A_\mu],\label{q1}\\
	A_\mu &= \{Q_{\bar c} ,\bar c\,A_\mu\},\label{q3}\\
		i D_\mu c &= [Q_{\rm BRST},  A_\mu]  ,\label{q2}\\
	\Delta_{Q_{\bar c}} &= \Delta_{D_\mu c}-d+1,\label{q4}
	\end{align}
where the first equation is \eqref{eq:VBRST}, and the last equation follows since $D_\mu c$ is the $\bar c$-shift current. Now we have:
\begin{align}
	\Delta_V &\stackrel{\eqref{q1}}{=} \Delta_{Q_\text{BRST}} - \Delta_{\bar{c}A}\nonumber\\
	&\stackrel{\eqref{q3}}{=}  \Delta_{Q_\text{BRST}} +\Delta_{A}- \Delta_{Q_{\bar{c}}} \nonumber\\
		&\stackrel{\eqref{q2}}{=} \Delta_{D_\mu c} - \Delta_{Q_{\bar{c}}}\nonumber\\
			&\stackrel{\eqref{q4}}{=} d-1\,.
\end{align}

The non-renormalization of the virial current operator in the Banks-Zaks fixed point was first addressed in \cite{Collins:1976yq} and reviewed in \cite{Nakayama:2016cyh} (see also \cite{Braun:2018mxm} in $4-\epsilon$ dimensions) based on the BRST analysis. We find our argument presented here much simpler and more transparent. The argument in this subsection applies to the Landau-gauge fixed point in the previous subsection, but we have presented it separately from the simpler argument available there.

\subsection{Crystalline membrane theory}

We now turn to crystalline membrane theory, which describes a $d$-dimensional membrane fluctuating around its equilibrium flat configuration in the ambient $D$-dimensional space. The most physically interesting case is $D=3$, $d=2$.
The Hamiltonian contains two fundamental fields $u_\mu$ and $h_a$, with indices $\mu = 1,\ldots,d$ parallel and $a=1,\ldots,D-d$ orthogonal to the membrane \cite{PhysRevLett.60.2634}
\begin{align}
\mathcal{H} = \frac{1}{2} \int d^d x \left[(\partial^2 h_a)^2 + \lambda (u_{\mu\mu})^2 + 2\mu u_{\mu\nu} u_{\mu\nu} \right] \, ,
\label{eq:cryst-memb}
\end{align}
where $u_{\mu\nu} = \frac12(\partial_\mu u_\nu + \partial_\nu u_\mu + \partial_\mu h_a \partial_\nu h_a)$.
If we set $h_a = 0$, the membrane model reduces to the theory of elasticity, a Gaussian theory which was observed to be scale invariant but not conformal invariant by Riva and Cardy \cite{Riva:2005gd}. Below we focus on the interacting case. In this case there is an IR fixed point at non-zero values of the couplings $\lambda$, $\mu$, which can be studied in a perturbative expansion in $d=4-\epsilon$. This is very interesting, as it implies that long-distance correlations of membranes are characterized by nontrivial critical exponents. Moreover, Mauri and Katsnelson \cite{Mauri:2021ili,Mauri-thesis}\footnote{See also these works for a thorough review of prior work on the membrane fixed point.} have recently shown that this fixed point is only scale invariant but not conformal. Therefore, it stands with the dipolar fixed point discussed by us as one of the only two currently known experimentally relevant non-Gaussian examples of scale without conformal invariance.

Note that model \eqref{eq:cryst-memb}, as all models described above, has a shift symmetry, which takes the form $u_\mu \to u_\mu + \epsilon_\mu$ with $\epsilon_\mu$ constant. The importance of this shift symmetry for controlling the renormalization structure of the model was already emphasized in \cite{Mauri:2021ili}. Ref.~\cite{Mauri:2021ili} also discussed the non-renormalization of the virial current dimension, and shift symmetry played a role in that discussion as well, along with other considerations. Here we wish to show that $\Delta_V=d-1$ can be understood in model \eqref{eq:cryst-memb}, as in all previously described models, as a \emph{direct} consequence of shift symmetry, via our general mechanism. We will see however that in this model the mechanism operates with a twist compared to the simple Eq.~\eqref{eq:Vcand}.

The first step of the argument is to express the trace of the stress tensor and to find a virial current. One finds that the stress tensor contains terms proportional to the beta-functions, which vanish at the fixed point, as well as a virial current term \cite{Mauri:2021ili}:
\begin{align}
\label{eq:virial_membrane}
 T_{\mu\mu} \big|_{\text{fixed point}}
 = - \partial_\mu V_\mu \, , \qquad
 V_\mu
 = V_\mu^{(0)}+ \partial_\nu \Om_{\mu\nu},\qquad V_\mu^{(0)} = k_1 u_\mu u_{\nu\nu}
 + k_2 u_\nu u_{\mu\nu} \, ,
\end{align}
where the precise form of the improvable part $\partial_\nu \Om_{\mu\nu}$, and the values of the constants $k_1, k_2$ will not be important for us.

Next we would like to connect the unimprovable virial current $V^{(0)}_\mu$ to the shift symmetry current, which is given by the expression 
\begin{equation}
	J_{\mu\nu} = \lambda u_{\alpha \alpha}\delta_{\mu\nu}  + 2\mu u_{\mu\nu}\,.
	\label{eq:Jmn}
	\end{equation}
As in the dipolar model, the shift current in the membrane model acquires a non-trivial scaling dimension $\Delta_J$. We can consider the trace part of the shift current and the traceless symmetric part. Although naively they do not mix under RG, they both should have the same scaling dimension, as a consequence of conservation of $J$.\footnote{This is analogous to the trace and the symmetric traceless part of the stress tensor having the same scaling dimension at the dipolar fixed point, see the discussion in Appendix \ref{sec:scaling}.} We conclude that 
\begin{equation}
	\label{eq:2fields}
	\Delta(u_{\alpha\alpha})= 	\Delta(u_{\mu\nu}-\text{trace}) = \Delta_J\,.
\end{equation}
To run the general argument for $\Delta_{V^{(0)}}=d-1$, we consider $[Q_\nu, V^{(0)}_\mu]$. In Eq.~\eqref{eq:Vcand}, this was equal to the shift current itself. In the membrane model, because of the coefficients $k_1,k_2$, this is a linear combination of two fields in \eqref{eq:2fields} which however both have the same dimension as the shift current (this is the twist alluded to above). Therefore the algebra works out the same, and we conclude that $\Delta_{V^{(0)}}=d-1$.

\subsection{Gaussian curvature interaction model}
\label{sec:mem-gauss}

Finally, let us consider the Gaussian curvature interaction (GCI) model \cite{Mauri:2020aga}, which provides an alternative description of crystalline membranes.
This model is obtained in $d=2$ by integrating out $u_\mu$ in \eqref{eq:cryst-memb}, and decoupling the resulting non-local interactions by introducing a Hubbard-Stratonovich field $\chi$.
The resulting effective Hamiltonian can be continued to arbitrary $d$, and reads
\begin{align}
	\label{eq:GCI}
\mathcal{H} = \int d^d x \left( \frac{1}{2} (\partial^2 h_a)^2 + \frac{1}{2v}  (\partial^2 \chi)^2  + \frac{i}{2} \chi (\partial^2 h_a \partial^2 h_a - \partial_\mu \partial_\nu h_a \partial_\mu \partial_\nu h_a) \right) \, ,
\end{align}
with $v$ the coupling constant.
The Hamiltonian for $v=0$ reduces to two copies of biharmonic theory, and it is thus conformal \cite{Nakayama:2019xzz}.
Instead, when $v \ne 0$ the coupling flows to a fixed point. This fixed point for $d=2$ is equivalent to the one considered in the previous subsection, but for generic $d$ it is distinct. Ref.~\cite{Mauri:2021ili} proved that this new fixed point also realizes scale without conformal invariance in any $d$. They also discussed why the virial current dimension does not get renormalized. 

The GCI model has a shift symmetry acting on the field $\chi$ as
\begin{equation}
	\chi\to \chi + a+ b_\mu x_\mu\,,
\end{equation}
where $a$ is a constant scalar and $b_\mu$ is a constant vector. We will refer to these as a ``constant shift symmetry'' and ``linear shift symmetry''.\footnote{We borrow this terminology from \cite{Griffin:2014bta}. This is also known as Galilean symmetry \cite{Nicolis:2008in} or dipolar global symmetry \cite{Gorantla:2022eem}.} The linear shift symmetry is a new feature of the model \eqref{eq:GCI} which was not present in other models discussed above.\footnote{In $d=2$, when the crystalline membrane model is equivalent to the GCI model, the linear shift symmetry in this subsection and the shift symmetry in the previous subsection are directly related.} This shift symmetry played a role, indirectly, in the discussion of the non-renormalization of $\Delta_V$ in \cite{Mauri:2021ili}, as they connected it to improved UV properties of the model, so that certain loop diagrams were finite.

Here we would also like to connect the non-renormalization of $\Delta_V$ to the linear shift symmetry. However, unlike \cite{Mauri:2021ili}, we would like to give an algebraic argument in the spirit of Eq.~\eqref{eq:Vcand}.

The virial current of the model has the form
\begin{equation}
	V_\mu = k_1 (\partial_\mu \chi) (\partial_\nu h_a \partial_\nu h_a) + k_2 (\partial_\nu \chi)(\partial_\nu h_a \partial_\mu h_a)+\partial_\nu \mathcal{O}_{\mu\nu}\,.
\end{equation}
The $k_1$ and $k_2$ are determined at the fixed point after renormalization (see \cite{Mauri:2021ili} for the renormalized expression, but we do not need them in the following). 
The $\mathcal{O}_{\mu\nu}$ in the improvable part of the virial current is given by
\begin{align}
	 \mathcal{O}_{\mu\nu} &=	k_3\,\partial_\mu \chi \partial_\nu \chi+ k_4  \delta_{\mu\nu} (\partial_\sigma \chi \partial_\sigma \chi)\,\nonumber\\
&+r_1 \delta_{\mu\nu} \partial^2 \chi + r_2  \partial_{\mu}\partial_{\nu}\chi + r_3 \partial_\mu h_a \partial_\nu h_a + r_4 \delta_{\mu\nu} (\partial_\sigma h_a \partial_\sigma h_a )\,.
 \label{eq:GCIimp}
	\end{align}
Without further conditions, all coefficients here are arbitrary. 

Under the linear shift symmetry, the virial current changes by $b_\nu[Q_\nu, V_\mu]$, where
\begin{align}
	[Q_\nu, V_\mu] = k_1 \delta_{\mu\nu} (\partial_\nu h_a \partial_\nu h_a) + k_2 (\partial_\nu h_a \partial_\mu h_a) + k_3 (\delta_{\mu\nu} \partial^2 \chi + \partial_\mu \partial_\nu \chi) + 2k_4  \partial_\mu \partial_\nu \chi . \label{v1}
\end{align}
Note that $\partial_\nu\mathcal{O}_{\mu\nu}$ terms corresponding to the second line of \eqref{eq:GCIimp} are linear shift-invariant and don't contribute.

We would like to relate the pieces in the r.h.s.~of this equation to the pieces of the linear shift symmetry current. The latter is given by
\begin{equation}
	J_{\mu\nu} = x_\nu \partial_\rho K_{\mu\rho} - K_{\mu\nu}\,,
\end{equation}
where $Q_\nu = \int d\Sigma^\mu J_{\mu\nu}$ is the corresponding charge. The symmetric 2-tensor field $K_{\mu\nu}$ enters the EOM for $\chi$, which can be written as 
\begin{equation}
	\label{eq:Kcons}
	\partial_\mu \partial_\nu K_{\mu\nu} = 0\,.
	\end{equation}
This implies the conservation of $J_{\mu\nu}$.

The	explicit form of the field $K_{\mu\nu}$, satisfying the ``partial conservation law''\footnote{Using the terminology of \cite{Dolan:2001ih}.} \eqref{eq:Kcons} is (classically)
\begin{align}
	K_{\mu\nu} = \frac{1}{v} \partial_\mu \partial_\nu \chi  - \frac{i}{2} (\delta_{\mu\nu} \partial_\rho h_a \partial_\rho h_a - \partial_\mu h_a \partial_\nu h_a ) + A (\partial_\mu \partial_\nu \chi - \delta_{\mu\nu} \partial^2 \chi) \ .  
	\label{k1}
\end{align}
Here the coefficient $A$ of the ``improvement term" is arbitrary. We wish to fix it so that $K_{\mu\nu}$ is a good scaling operator. In the following, we do not need to know the explicit value of $A$.
Note that, like for the shift current \eqref{eq:Jmn} from the previous section, the partial conservation of $K_{\mu\nu}$ implies that the whole of $K_{\mu\nu}$ will have the same scaling dimension, i.e. the trace part $\Km=K_{\mu\mu}$ and the traceless symmetric part $\mathcal{K}_{\mu\nu}$ of $K_{\mu\nu}$ have the same dimension $\Delta_K$.

We now consider vector operators on the first level of the linear shift symmetry multiplets built on top of $\mathcal{K}$ and $\mathcal{K}_{\mu\nu}$. These are defined as the operators $\mathcal{V}_{1,\mu}$ and $\mathcal{V}_{2,\mu}$ which have a well-defined scaling dimension and satisfy the equations:
\begin{align}
	[Q_\nu, \mathcal{V}_{1,\mu}]=\delta_{\mu\nu}\mathcal{K}\,,\qquad
	[Q_\nu, \mathcal{V}_{2,\mu}]=\mathcal{K}_{\mu\nu}\,.
	\end{align} 
Since $\mathcal{V}_{1,\mu}$ and $\mathcal{V}_{2,\mu}$ are scaling operators, we conclude, by Eq.~\eqref{eq:Vcand}, that they both have dimension $d-1$. Thus any linear combination
\begin{equation}
	\mathcal{V}_{\mu} = p_1 \mathcal{V}_{1,\mu} + p_2 \mathcal{V}_{2,\mu}
	\end{equation}
is a candidate virial current.

What is the relation of this construction to the true virial current $V_\mu$ given above, which transforms under $Q_\nu$ as \eqref{v1}? Let us choose the constants $p_1,p_2$ in $\mathcal{V}_{\mu}$ and the improvable terms $k_3,k_4$ in $V_\mu$ so that
\begin{equation}
	\label{QV}
	[Q_{\nu}, V_\mu -  \mathcal{V}_\mu] = 0\,.
\end{equation} 
To achieve this, we first determine $p_1$ and $p_2$ from $k_1$ and $k_2$ by comparing the $h$ dependent terms of \eqref{k1} and \eqref{v1}. Then, we fix the shift non-invariant improvable terms in $V_\mu$ (i.e. $k_3$ and $k_4$) by comparing the $\chi$ dependent terms. The shift-invariant part of $V_\mu$ is left undetermined. 

Now, Eq.~\eqref{QV} shows that the difference $v_\mu := V_\mu-\mathcal{V}_\mu$ is linear shift invariant. Inspecting all vector operators of the appropriate classical scaling dimension, of schematic form $\partial^3\chi$, $\partial \chi \partial^2\chi$ and $\partial h\partial^2 h$, it turns out that all such terms are improvable \cite[Eq.~(B.2)]{Mauri:2021ili}, namely~of the form $\partial_\nu \mathcal{O}_{\mu\nu}$ with $\mathcal{O}_{\mu\nu}$ in the second line of \eqref{eq:GCIimp}. Hence, we can improve the stress tensor, so that the improved virial current is $V'_\mu = V_\mu-v_\mu$. By construction, this final virial current satisfies 
$V_\mu'=\mathcal{V}_\mu$, and hence $\Delta_{V'}=d-1$, completing the argument.

\subsection{Higher derivative shift symmetric scalar}
So far in this section, we have discussed scale invariant but non-conformal theories proposed in the literature, verifying the non-renormalization of the virial current.
As a further application with novel predictions, let us study an interacting theory of higher derivative shift symmetric scalar.  We consider the action studied in \cite{Safari:2021ocb}\footnote{
	A multi-component generalization of this model may be related to the membrane theories discussed above. See \cite{Delzescaux:2023rgm} for more details. The following discussion applies to their models, too.}
\begin{align}
S = \int d^dx \left(\frac{1}{2} (\partial^2 \varphi)^2 + g (\partial_\mu \varphi \partial_\mu \varphi)^2 \right) \ .
\end{align}
It is invariant under the constant shift $\varphi \to \varphi + c$. In $d=4$, the action is conformal invariant classically (broken by the RG effect),  and in any dimensions it is conformal invariant at the non-interacting fixed point $g_*=0$.  It is interesting to see if the interacting fixed point in $d=4-\epsilon$ dimension is scale-invariant or conformal invariant. Note that the theory is non-unitary.

The existence of a non-trivial fixed point was confirmed by the perturbative calculation of \cite{Safari:2021ocb}, which is located at $g_* = O(\varepsilon)$ at one-loop. The scaling dimension of $\varphi$ is $\Delta_{\varphi} = \frac{d-4}{2} +\eta$, where the anomalous dimension starts at three loops: $\eta=\frac{1}{25}\varepsilon^3+O(\varepsilon^4)$ \cite[Eq.~(III.47)]{Safari:2021ocb}. Is this interacting fixed point conformal invariant?

One of the results in \cite{Safari:2021ocb} was that the fixed point is conformal up to one loop in perturbation theory. We will show here that it is only scale invariant if higher orders are taken into account.
The crucial observation is that this theory has a shift symmetry generated by the conserved shift current $J_\mu = \partial_\mu \partial^2 \varphi -4 g \partial_\mu\varphi \partial_\nu\varphi \partial_\nu \varphi$. Note that unless $g = 0$ at the fixed point, the entire $J_\mu$ cannot be written as a derivative of other local operators. From our general argument of the shift symmetry, the dimension of $J_\mu$ satisfies $\Delta_{J} + \Delta_{\varphi} = d-1$, which implies $\Delta_{J} = \frac{d+2}{2} - \eta $. The scaling dimension $\Delta_{J}$ is {\it not} $d-1$ (unless $d=4$), violating a necessary condition for a conserved {\it primary} current in conformal field theories. We conclude that the interacting fixed point cannot be conformal within perturbation theory.

Regarding the trace of the stress tensor, a calculation analogous to Section \ref{sec:stress-tens-arg} shows there is a virial current given by $V_\mu = -\Delta_\varphi\, \varphi J_\mu$. Thanks to the shift symmetry, our general argument shows that the scaling dimension of $V_\mu$ is protected to be $d-1$ exactly, while this operator is not conserved at the interacting fixed point. A general lesson here is that at interacting fixed points with shift symmetry, it is more natural that they are only scale invariant rather than conformal invariant.

\section{Conclusions}
\label{sec:conclusion}

In this paper we discussed a fascinating RG fixed point which deserves to be more widely known - the dipolar fixed point of Aharony and Fisher, describing the phase transition in isotropic ferromagnets with strong dipole-dipole forces. Our interest in this fixed point was sparked by the realization that it provides an example of an interacting theory that is scale but not conformally invariant. Such examples are rare \cite{Nakayama:2016cyh}, and experimentally relevant ones are even rarer, the only other one occurring in the physics of fluctuating membranes \cite{Mauri:2021ili}.

One of the most pleasing conclusions of our work is a new insight into the role of a shift symmetry in protecting the virial current dimension from loop corrections due to interactions. Since the virial current $V_i$ is mapped by the shift symmetry charge into the shift symmetry current, we naturally obtain $\Delta_V=d-1$.
Furthermore, by going through the list of other known interacting scale without conformal models, we found that all of them have a shift symmetry and protect the virial current dimension via the same mechanism or its small variation. While we do not have a proof, could it be that shift symmetry is a necessary feature of such models?

The shift symmetry is always spontaneously broken in the sense that there exists an operator $\mathcal{O}$, whose variation is a constant (so that $\langle \delta \mathcal{O} \rangle \neq 0$). Usually, we expect that a spontaneously broken global symmetry leads to a massless Nambu-Goldstone boson particle, but in our examples, this does not happen. Instead, the infrared theory is a scale-invariant fixed point that is non-trivially interacting, with anomalous dimensions. It is instructive to understand how this is avoided. The key point is that this may only happen in a free theory or in a non-unitary model. Indeed all our examples were non-unitary.

In unitary (relativistic) quantum field theories, the momentum space 2pt function of the spontaneously broken current $\langle J_i(q) \mathcal{O} (-q) \rangle = i\frac{q_i}{q^2}$ implies the existence of a massless Nambu-Goldstone boson by inserting the complete momentum eigenstates. In particular, we then predict the existence of a $1/q^2$ pole in the $\langle \mathcal{O}(q) \mathcal{O}(-q) \rangle$ 2pt function. This means that the IR theory is that of a free massless boson (which indeed has a shift symmetry).

On the contrary, the dipolar fixed point, where $\mathcal{O}$ here is given by $U$, avoids the existence of $1/q^2$ pole in $\langle \mathcal{O}(q) \mathcal{O}(-q) \rangle$. It is allowed to do so because it's non-unitary. Technically, the Lorentzian continuation should give rise to the structure of a Hilbert space with an indefinite metric, i.e.~$1=\sum_n |n \rangle \langle n|$ is replaced with  $1=\sum_n |n \rangle \eta^{nm} \langle m|$, where $\eta^{nm}$ is not positive definite. The appearance of the indefinite metric $\eta^{nm}$ avoids the usual argument.

It is still an open question if scale invariance implies conformal invariance in {\it interacting and unitary} quantum field theories in three dimensions.\footnote{In 4d,  Refs.~\cite{Dymarsky:2013pqa,Dymarsky:2014zja} proved that a unitary scale-invariant theory \emph{without dimension 2 scalars} must be conformal. With dimension 2 scalars, they showed that the theory is either conformal, or the trace of the stress tensor must have the form $T_{\mu\mu}=\partial^2 \mathcal{A}+\mathcal{B}$ where $\mathcal{A}$ is a dimension 2 scalar, and $\mathcal{B}$ is a generalized free field of dimension 4. The latter loophole is still open to the best of our knowledge.} 
As argued above, such theories cannot have a shift symmetry, so if they exist, the virial current dimension should be protected by another mechanism.

Going back to the dipolar fixed point, although some of its features were observed (as we reviewed in Section \ref{sec:exp}), more experimental studies are welcome. The dipolar critical exponents are relatively poorly known compared to the Heisenberg fixed point. Perturbative results are available only at three loops \cite{Kudlis:2022rmt}. On the nonperturbative side, we mention preliminary computations of critical exponents using the functional RG \cite{ERG-talk}. Since the fixed point is not conformal, the conformal bootstrap \cite{Poland:2018epd} does not apply. This is then a good concrete model to think about developing bootstrap techniques in the absence of conformal invariance. Indeed, we still have the operator product expansion (OPE). In addition to scale invariance, the model possesses a shift symmetry. The shift multiplets, a notion which we introduced in Section \ref{sec:shift}, have a particular structure, and may play a role similar to the conformal multiplets in setting up the bootstrap calculation.  It is important to ascertain if different operators in the same shift multiplets have their OPE coefficients related. A further hurdle is the lack of unitary, hence an analog of Gliozzi's method \cite{Gliozzi:2013ysa} will be called for, instead of techniques based on positivity \cite{Rattazzi:2008pe}. This is a hard but very interesting problem.

\acknowledgments

SR thanks Viacheslav Krivorol for discussions related to Section \ref{sec:exp}. We thank Eric Perlmutter for a prescient question about an axiomatic definition of shift symmetry.
SR and AGG are supported by the Simons Foundation grant 733758 (Simons Bootstrap Collaboration) and AGG is also supported by the Simons Foundation grant 915279 (IHES).
The work by YN is in part supported by JSPS KAKENHI Grant Number 21K03581. 
\appendix

\section{Demagnetizing factor}
\label{app:demag}

In this appendix we explain Eq.~\eqref{eq:demag} in the main text, needed to
interpret all experimental papers measuring magnetic susceptibility. We
normalize $\phi$ as in \eqref{normM}, i.e. $\phi = M$.

When a sample is put in an external magnetic field $B^{(0)}_i$, which we
assume uniform, it gets magnetized. Magnetization inside the sample $\phi_i
(x)$, $x \in \Omega$, can be found by minimizing the Hamiltonian:
\begin{equation}
  \int_{\Omega} \left(\frac{1}{2} a (\partial \phi_i)^2 + \frac{1}{2} b \phi_i^2 -
  \phi_i B^{(0)}_i \right) + \frac{1}{2}  \iint_{x, y \in \Omega}
  U_{ij}  (x - y) \phi_i (x) \phi_j (y), \label{eq:eff-phi}
\end{equation}
where the integration is over the sample $\Omega$. Suppose we are above $T_c$,
then $b > 0$ and the quadratic form depending on $\phi_i$ is positive
definite, so the minimizer (with Neumann boundary conditions) exists and is
unique. It solves the classical equation of motion:
\begin{equation}
  - a \partial^2 \phi_i + b \phi_i - B^{(0)}_i + \int_{y \in \Omega} U_{ij}
  (x - y) \phi_j (y) = 0 . \label{magn-eom}
\end{equation}
The most important particular case arises when the magnetization is constant,
in which case the first term in \eqref{magn-eom} drops out. There is a
consistency condition for this to happen: the integral
\begin{equation}
  \int_{y \in \Omega} U_{ij}  (x - y) {= 4 \pi D_{i j}}  \qquad (x \in \Omega)
  \label{eq:dem1}
\end{equation}
must be $x$-independent. Famously, this happens if the sample is ellipsoidal;
the tensor $D_{i j}$ is then diagonal $D_{i j} = \delta_{i j} D_i$ in the
ellipsoid axis. The factors $D_i$ are called demagnetizing factors. For an
ellipsoid $\frac{x_1^2}{a_1^2} + \frac{x_2^2}{a_2^2} + \frac{x_3^2}{a_3^2}
\leqslant 1$ they are given by (see e.g. section 4.18 in
{\cite{stratton2007electromagnetic}}, the solution relies on ellipsoidal
coordinates which are nicely reviewed in {\textsection}4 of
{\cite{landau2013electrodynamics}}):
\begin{equation}
  D_i = \frac{a_1 a_2 a_3}{2}  \int_0^{\infty} \frac{d s}{(s + a_i^2)
  \sqrt{(s + a_1^2)  (s + a_2^2)  (s + a_3^2)} \hspace{0.17em}}
  \hspace{0.17em} . \label{eq:demag1}
\end{equation}
In older literature {\cite{stoner1945xcvii,osborn1945demagnetizing}}, the
demagnetizing factors were expressed in terms of elliptic integrals and
tabulated. However, for practical purposes it is easier nowadays to
numerically integrate the definition in \eqref{eq:demag1}. In general we have
a relation $D_1 + D_2 + D_3 = 1$ {\cite{osborn1945demagnetizing}}, in
particular $D  = 1 / 3$ for the sphere.

It follows from the above discussion that for ellipsoids the magnetization is
constant and is given by Eq.~\eqref{eq:demag}. If the shape is not ellipsoidal
then one has to solve Eq.~\eqref{magn-eom}, including the first term, so the
discussion becomes more complicated. In experiments one usually uses spherical
samples. Sometimes cylindrical samples are used, but instead of solving Eq.
\eqref{magn-eom} one approximates them by ellipsoids and pretends that the
magnetization is constant.

In textbooks, magnetization phenomena are usually discussed in terms of the
$H$ field. Let us see how this is related to the above. Let $B _i$ be the
magnetic field produced by the magnetization $\phi_i$, and $ B^{(0)}  + B $ be
the total field. We can consider the Hamiltonian in which the magnetic field
$B$ is explicit (see Eq.~\eqref{eq:HB}, where we have to rescale $\phi_i$ to set $z=1$, and drop the quartic term):
\begin{equation}
  \int_\Omega \left( \frac{1}{2} a (\partial \phi )^2 + \frac{1}{2} \tilde{b} \phi^2 - \phi
  (B^{(0)} + B) + \frac{1}{8 \pi} B^2 \right), \label{eq:Bexpl}
\end{equation}
where $B = \nabla \times A$. Varying this Hamiltonian over $A$ we get the
equation
\begin{equation}
  \nabla \times (B - 4 \pi \phi) = 0, \label{eq:Heq}
\end{equation}
which is usually written as $\nabla \times H = 0$, introducing the field $H =
B - 4 \pi \phi$.

We can also integrate out $B$ from \eqref{eq:Bexpl}, and get an effective
Hamiltonian just in terms of the $\phi$ field. It is easy to show that this
Hamiltonian takes the form \eqref{eq:eff-phi} with an important mass shift (see Eq.~\eqref{shift} where we need to put $z=1$)
\begin{equation}
  b = \tilde{b} - 4 \pi . \label{eq:shift}
\end{equation}
The equation for $H$ is usually solved by introducing magnetic potential $U$
so that $H = - \nabla U$. The condition that $B$ is solenoidal then gives
\begin{equation}
  \nabla^2 U = 4 \pi \nabla \cdot \phi,
\end{equation}
and see e.g. {\cite{jackson1977classical}} for appropriate boundary conditions
that $U$ must satisfy on the boundary of the sample. For constant
magnetization $\phi$, one finds
\begin{equation}
  U (x) = \int_{\partial \Omega} d^2 x' \frac{n' \cdot \phi}{| x - x' |},
\end{equation}
which gives the $H$ field inside the sample:
\begin{equation}
  H (x) = - 4 \pi D_{i j} (x) \phi_j, \qquad x \in \Omega,
\end{equation}
\begin{equation}
  D_{i j} (x) = \frac{1}{4 \pi} \int_{\partial \Omega} d^2 x' \partial_{x_j}
  \frac{1}{| x - x' |} n'_j = \frac{1}{4 \pi} \int_{\Omega} d^3 x' U_{i j} (x
  - x'),
\end{equation}
by Stokes' theorem. As we already discussed for the ellipsoidal sample the
last integral does not depend on $x$ and agrees with the demagnetizing factor
defined in \eqref{eq:dem1}.

Finally, let us derive Eq.~\eqref{eq:demag}. Extremizing the Hamiltonian
\eqref{eq:Bexpl} over $\phi$ we get, assuming $\phi$ and $B$ are constant
inside the sample:
\begin{equation}
  \tilde{b} \phi_i = B_i^{(0)} + B_i
\end{equation}
Substituting into this equation $B = H + 4 \pi \phi$ and $H_i = - 4 \pi D_i
\phi_i$, and using the mass shift relation \eqref{eq:shift}, we get precisely
\eqref{eq:demag}. Defining the total $H$ field $H^t = B^{(0)} + H$, Eq.
\eqref{eq:demag} can also be written as
\begin{equation}
  \phi  = b^{- 1} H^t . \label{eq:phiHt}
\end{equation}
\section{Experimental data}\label{app:experiment}

In this appendix we describe the experimental data used to extract Table
\ref{tab:exp-data}.

\paragraph{Amplitude $C$} This amplitude is measured from the static
susceptibility above the Curie temperature $\chi = \frac{\partial M}{\partial
H ^t} |_{H^t = 0}$. Here $H^t$ is the total $H$ field in the sample, i.e. $H^t
= H^{(0)} + H_{\text{}}$, where $H^{(0)} = B^{(0)}$ is the applied field, and
$H = - 4 \pi D M$ is the field associated with the magnetization of the
sample, see Appendix \ref{app:demag} for details. Susceptibility varies as
$\chi = Ct^{- \gamma}$ near the Curie point, which is consistent with
\eqref{Bt} using Eq.~\eqref{eq:phiHt}. In this paper we work in the Gaussian
units, and must keep careful track of factors of $4 \pi$ compared to the
literature.

EuS: We use {\cite{kotzler1986change}} for the susceptibility near $T_c =
16.56$ K. They report their results in terms of normalized magnetization $m =
M / M_0$ and applied field $h = H / H_c$. Thus, our amplitude $C$ is related
to the value $\Gamma$ they report as $C_{\rm{EuS}} = \Gamma M_0 / H_c$, with the measured values $(\Gamma, 4 \pi M_0, H_c)_{\rm{EuS}} = (0.45, 15.4 \hspace{0.17em} \text{kOe}, 35.4 \hspace{0.17em} \text{kOe})$,\footnote{One must be careful because $M^{\rm{there}} = 4 \pi M^{\text{here}}$, because
they write $H = H_{\rm{ext}} - NM$ and their demagnetizing factor for a
sphere is $N=1/3$.} which gives $C_{\text{EuS}} \approx 0.016$. Note that
we extract $\Gamma$ from their Fig.~3 as the point where the upper curve
intersects the $y$-axis, so $\Gamma$ is slightly larger than what they call
$\Gamma_0$. (Same $\Gamma$ can also be extracted from the right part (small
$m$) of the lower curve in their Fig.1.)

EuO: Similarly to EuS ($T_c = 69.1$K), the measures for EuO are $(\Gamma, 4
\pi M_0, H_c)_{\text{EuO}} = (0.397, 24 \hspace{0.17em} \text{kOe}, 146
\hspace{0.17em} \text{kOe})$ {\cite{PhysRevB.12.5255}}, giving $C_{\text{EuO}}
\approx 0.052$.

Fe: In {\cite{NoakesEtAl66}}, the susceptibility for iron is reported as $\chi
= T_1^{\gamma}  (T - T_c)^{- \gamma}$ so $C_{\text{Fe}} = (T_1 /
T_c)^{\gamma}$ with $(T_1, T_c, \gamma) = (1.27 \hspace{0.17em} \text{K}, 1043
\hspace{0.17em} \text{K}, 1.333)$.

Ni: We use a recent comprehensive Ref. {\cite{SeegerEtAl95}}. Since they work
in SI units, the susceptibility is $\chi^{\text{there}} = 4 \pi
\chi^{\text{here}}$, so the amplitude is $C_{\text{Ni}} = J_0 / (4 \pi h_0)$.
They fit the data in two different ways, and we take the best which includes
corrections to scaling (CTS). We take $h_0 / J_0 = 1991.5$, the average over
the four samples of different shapes.

\paragraph{Amplitude $f^+$} This amplitude is measured from the divergence of
the correlation length $\xi = f^+ t^{- \nu}$ in neutron scattering experiments
above the Curie temperature. Refence {\cite{PhysRevB.14.4908}} considers both
EuS and EuO, and in their notation $f^+ = a_{nn} / F^+$ with $(F^+,
a_{nn})_{\text{EuS}} = (2.33, 4.22 \hspace{0.17em}${\r A}$)$ and $(F^+,
a_{nn})_{\text{EuO}} = (2.32, 3.64 \hspace{0.17em}${\r A}$)$. For iron, the
amplitude is reported in {\cite{BALLY1968396}} as $f^+ = 1 / A$ with $A = 1.1\text{\r A}{}^{- 1}$. This value also fits reasonably well the
data in table I of {\cite{PhysRev.139.A1866}} and in table V of
{\cite{PhysRev.142.291}}. For nickel, we use ${f^+}^{}$ reported in
{\cite{Ni-xi}} (called $\xi_0$ there).

\section{Microscopic model}
\label{app:micro}

This appendix estimates the parameters $a$ and $b$ in the Hamiltonian \eqref{Hpheno} starting from a microscopic model of dipoles.
After presenting the model, we study it in mean-field theory using a Hubbard-Stratonovich transformation.
We conclude by comparing the results for EuS and EuO to the experimental Table \ref{tab:exp-data}.

\subsection{Model}

Consider a system of atoms in a cubic lattice of type sc, bcc or fcc.
The atom at point $x$ has dipole moment $\vec m=(m^i)_{i=1,2,3}$ of magnitude $|\vec m| = \mu$.
The dipoles interact with a dynamical magnetic field $\vec B$, so the partition function is
\begin{align}
 \label{eq:Zmodel}
 Z = \int D\vec B \, \prod_{x} d^3m_x \,
     \delta\!\left(\vec m_x^2 - \mu^2 \right)
     \exp \big( {-}\beta \Hm[m,B] \big) \, .
\end{align}
Recall that the energy felt by a dipole in a magnetic field is $- \vec m \cdot \vec B$.
At site $x$ the total magnetic field is $\vec B^t_x \equiv \vec B^{(0)}(x) + \vec B(x)$, where $\vec B^{(0)}$ is a background field and $\vec B$ is dynamical.
In total, the electromagnetic part of the Hamiltonian is
\begin{align}
\label{eq:ham-EM}
 \Hm_{\text{EM}}
 = - \sum_{x} \vec m_x \cdot \vec B^t_x
 + \frac{1}{8\pi} \int d^3x \, \vec B(x)^2 \, .
\end{align}
Besides electromagnetic interactions, the dipoles experience short-range ferromagnetic interactions, which can be modeled with the Hamiltonian
\begin{align}
\label{eq:ham-sr}
 \Hm_{\text{short-range}}
 = \frac{J}{4c} \sum_{x,\delta} (\vec m_x - \vec m_{x+\delta})^2
 - J \theta \sum_x \vec m_x^2 \, .
\end{align}
Here $J$ is the interaction strength, $\delta$ runs over nearest neighbors, and $c$ is the number of nearest neighbors.
Because of the integration measure \eqref{eq:Zmodel}, the term proportional to $\theta$ only changes the partition function by an overall normalization.
We then expect that sensible predictions of our model should be independent of $\theta$.
We choose $\theta > 1$, which ensures that if we rewrite the interactions as
\begin{align}
\label{eq:ham-sr-K}
 \Hm_{\text{short-range}}
 = - \frac12 \sum_{x,y} \vec m_x K_{xy} \vec m_{y} \, ,
\end{align}
then the quadratic form $K_{xy}$ is positive definite, a property necessary for the Hubbard-Stratonovich transformation.

\subsection{Hubbard-Stratonovich transformation}
\label{sec:HS}

To compare with the Hamiltonian \eqref{Hpheno}, we need to express the partition function in terms of a coarse-grained magnetization that is not restricted by $|\vec m| = \mu$.
This is achieved with a Hubbard-Stratonovich transformation\footnote{See e.g. \cite{FisherNotes} for a detailed introduction.}
\begin{align}
 Z = \int \prod_{x} d^3 \lambda_x \, d^3m_x \,
     \delta\!\left( m_x^2 - \mu^2 \right)
     \exp \left(
       - \frac \beta2 \sum_{x,y}
         \vec \lambda_{x} K^{-1}_{xy} \vec \lambda_{y}
       + \beta \sum_{x} \vec m_x \cdot (\vec \lambda_x + \vec B^t_x)
     \right) \, .
\end{align}
For clarity, we ignore terms that only change the normalization of the partition function. We also omit the action for the dynamical magnetic field $\vec B$, but we shall restore it at the end.
Now we integrate over $\vec m$ using
\begin{align}
 \frac{1}{4\pi}
 \int d^3m \,
 \delta\!\left( \vec m^2 - 1 \right)
 \exp \big(\vec m \cdot \vec v \, \big)
 = \frac{\sinh |\vec v|}{|\vec v|}
 = \exp \left( \frac{1}{6} \vec v^2 -\frac{1}{180} (\vec v^2)^2 + \ldots \right) \, ,
 \label{eq:mint}
\end{align}
where $\vec v = \beta \mu (\vec \lambda + \vec B^t)$.
Since we are interested in a mean-field theory analysis, we drop quartic powers of $\vec v$ and higher, so
\begin{align}
 \label{eq:ham-lam}
 \Hm
 = \frac 12 \sum_{x,y} \vec \lambda_{x} K^{-1}_{xy} \vec \lambda_{y}
 - \frac{\beta\mu^2}{6} \sum_x (\vec \lambda_x + \vec B_x^{t})^2 \, .
\end{align}
This depends on the field $\vec\lambda$ that takes arbitrary real values, so we can interpret $\vec \lambda$ as a coarse-grained field related to the magnetization, with the precise relation given below.
That the last term in \eqref{eq:ham-lam} comes with a negative coefficient is physically reasonable---lowering the temperature should have an ordering effect.

Now we should evaluate the inverse of the quadratic form $K$.
Before calculating the inverse, note that for $\vec \lambda$ varying slowly compared to the lattice size, $K$ acts as\footnote{We use the relation $\sum_{\delta} (q \cdot \delta)(p \cdot \delta) = \frac{c}{3} \, \textrm{a}^2 \, q \cdot p$, which is valid for cubic lattices. Here $c$ is the coordination number of the lattice, and the distance to nearest neighbors is $|\delta| = \textrm{a}$.}
\begin{align}
 \sum_{x,y} \vec \lambda_x K_{xy} \vec \lambda_{y}
 =
 \int d^3x \left(
 \frac{2 J \theta}{V} \vec \lambda^2
 - \frac{J\textrm{a}^2}{6V} (\partial_i \vec \lambda)^2
 + O(\partial^4 \lambda^2)
 \right) \, .
 \label{eq:Ksmooth}
\end{align}
In the previous equation, $\textrm{a}$ is the nearest-neighbor distance and $V$ is the volume of the unit cell, which appears in the continuum limit $\sum_x \to \int \frac{d^3x}{V}$.
Now we can invert $K$ using \eqref{eq:Ksmooth} and treating the kinetic term as a perturbation, so $(I - \varepsilon A)^{-1} \approx I + \varepsilon A$. 
This approximation is valid for sufficiently slow fluctuations, or in other words, the higher-order terms are irrelevant in the RG sense.
Combining all the ingredients, we arrive at
\begin{align}
 \label{eq:ham-phi-B}
 \Hm
 = \int d^3 x \left(
 \frac12 c_1 (\partial_i \vec \lambda)^2
 + \frac12 c_2 \vec \lambda^2
 - \frac{1}{2} c_3 \big( \vec \lambda + \vec B^t \big)^2
 \right)
 \, ,
\end{align}
where $c_{i}$ are given by
\begin{equation}
 c_1 = \frac{\mathrm{a}^2}{24 \theta ^2 J V} \, , \qquad
 c_2 = \frac{1}{2 \theta  J V} \, , \qquad
 c_3 = \frac{\beta  \mu ^2}{3 V} \, .
\end{equation}
Now we redefine
\begin{align}
 \vec \lambda = p \vec B^t + q \vec \phi \, , \quad
 p = - 1 / (1 + \sqrt{{c_2}/{c_3}}) \, , \quad
 q =   1 / \sqrt{c_2 c_3} \, .
 \label{eq:lam2phi}
\end{align}
Here $p$ is chosen so that the term $(\vec B^{t})^2$ has zero coupling, and $q$ so that the linear term $-\vec B^{t} \cdot \vec \phi$ is unit normalized. This ensures that $\vec \phi$ is the correct coarse-grained magnetization, because it satisfies $\langle \vec \phi \rangle = \frac{1}{\beta} \frac{\delta \log Z}{\delta B^{(0)}}$. The action in terms of $\vec\phi$ takes the form
\begin{align}
	\Hm
	= \int d^3 x \left(
	\frac{c_1}{2 c_2 c_3} (\partial_i \vec \phi)^2
	+ \frac{c_2 - c_3}{2 c_2 c_3} \, \vec \phi^{\,2}
	- \vec B^{t} \cdot \vec \phi
	\right)\, .
\end{align}
Here we ignored terms $\partial_i \vec B^t=\partial_i \vec B^{(0)}+\partial_i \vec B$, justified for a sufficiently homogeneous external field and, for $\partial_i \vec B$, because upon integrating out the magnetic field $\vec B$, these terms will generate higher derivative interactions, which are more irrelevant in the RG sense.

Finally, we add the action $\frac{1}{8\pi} \int \vec B^2$ for the dynamical magnetic field and integrate it out, as explained in Appendix \ref{app:demag}. This generates the long-range term $\phi.U.\phi$, and adds $-4\pi$ correction to the mass.
The effective Hamiltonian comes out to be
\begin{align}
 \label{eq:ham-phi}
 \Hm
 = \int d^3 x \left(
 \frac12 a (\partial_i \vec \phi)^2
 + \frac12 b \vec \phi^{\,2}
 - \vec B^{(0)} \cdot \vec \phi
 \right)
 + \frac{1}{2} \int d^3x \, d^3y \, \phi^i(x) \phi^j(y) U_{ij}(x-y) \, ,
\end{align}
where
\begin{align}
 a = \frac{\text{a}^2 V}{4 \beta  \theta  \mu ^2} \, , \qquad
 b = \frac{3 V}{\beta  \mu ^2} - 2 \theta J V - 4 \pi \, .
 \label{eq:params-abe}
\end{align}
We see that our calculation gave an effective Hamiltonian of the same form as \eqref{Hpheno}, \eqref{eq:eff-phi} considered above.
Note that the coefficient $b\propto \beta^{-1}$ at $\beta\ll 1$. This is in agreement with Curie's law, that the susceptibility $\chi\propto 1/T$ at high temperatures.

The identification of the correct coarse-grained magnetization $\phi$ in terms of $\lambda$ was crucial in the above line of reasoning. The derivation is robust and would work for any quadratic Hamiltonian density of the form
$
k_1 (\partial \lambda)^2+ k_2 \lambda^2 +2 k_3 \lambda\cdot B^t +k_4 (B^t)^2
$
as long as the quadratic form $k_2 x^2 +2 k_3 x y+k_4 y^2$ is not sign-definite.

\subsection{Comparison to Europium compounds}

\begin{figure}[ht]
	\centering
	\includegraphics[scale=0.6]{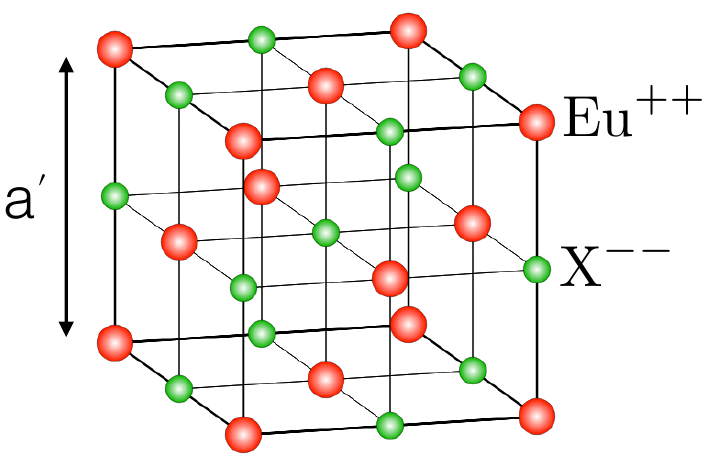}
	\caption{\label{fig-EuX}Fcc lattice structure of EuX, X=S,O.}
\end{figure}

We now apply these results to the ferromagnetic insulators EuS and EuO.
These compounds form an fcc lattice of the rock salt type, so the nearest-neighbor distance is related to the lattice constant $\textrm{a}'$ by $\textrm{a} = \textrm{a}' / \sqrt{2}$, and the volume of the unit cell is $V = \textrm{a}^3 / \sqrt{2}$ (Fig.~\ref{fig-EuX}).
As experimental inputs, we will use the lattice constant and the critical temperature of these materials:
\begin{align}
	\text{EuS}:\quad \textrm{a}' = 5.96 \,\textrm{\AA} \,, \quad T_c = 16 \, \mathrm{K},\nonumber\\
	\text{EuO}:\quad \textrm{a}' = 5.14 \,\textrm{\AA} \,, \quad T_c = 69 \, \mathrm{K}.\label{exp-EuX}
\end{align}

In the europium compounds, the Eu$^{2+}$ ions are responsible for the dipole interactions.
The europium ion has spin $S = 7/2$ so the Land\'e $g$-factor is $g=2$.
The magnitude of the dipole moment is then $\mu = g \mu_B \sqrt{S(S+1)}$, where $\mu_B$ is the Bohr magneton.

We start estimating the coupling $J$. For this we use that $b(T_c) = 0$, where
$b$ is given in \eqref{eq:params-abe}. We obtain:
\begin{align}
J = \frac{3 k_B T_c}{2 \theta \mu ^2} - \frac{2 \pi}{\theta V} \, .
\end{align}
As a side comment, note that we can introduce a parameter $\hat{g}$ that measures the relative shift of the critical temperature due to dipolar effects, namely:
\begin{align}
  \hat g
  = \frac{T_c - T_c^{\text{Heis.}}}{T_c}
  = \frac{4 \pi  \mu ^2}{3 k_B T_c V} \, ,
  \label{eq:ghat}
\end{align}
where $T_c^{\text{Heis.}}$ is defined as the temperature where $b^{\text{Heis.}}$ vanishes, which is defined by dropping in the expression for $b$ the $4\pi$ term, which is due to dipolar effects. It is in this meaning that quantity $\hat g$ was introduced by Aharony and Fisher \cite{FisherAharony}, who used it to estimate the size of dipolar effects. (Note that the expression for $\hat g$ in \cite{FisherAharony} is a factor of 3 smaller than ours, while another paper by the same authors \cite{AharonyFisher} leads to $\hat g$ that is a factor 3 larger than ours, see Remark \ref{AFremark} below.)
Using the experimental values \eqref{exp-EuX}, we evaluate
\begin{align}
	\hat g_{\text{EuS}}
	= 0.19 \, , \qquad
	\hat g_{\text{EuO}}
	= 0.07 \, .
\end{align}
	
Knowing $J$ we can find $a$ and $b$ in \eqref{eq:params-abe}. Expanding near $T_c$, we obtain
\begin{align}
 b = C^{- 1} t,\qquad	\xi=(a/b)^{1/2} = f^{+} t^{-1/2},
\end{align}
with the following expressions for the critical amplitudes
\begin{align}
 C = \frac{\mu ^2}{3 k_B T_c V} \, , \qquad
 f^+ = \frac{\textrm{a}}{2 \sqrt{3 \theta}} \, .
 \label{eq:critamp}
\end{align}
Recall that $\theta>1$ is an arbitrary parameter that we introduced to make the quadratic form in \eqref{eq:ham-sr-K} positive definite.
If we could compute the partition function exactly, then $\theta$ would only affect the overall normalization without changing physical observables.
The fact that $f^+$ depends on $\theta$ suggests that this prediction has to be taken with a grain of salt, unlike the predictions for $\hat g$ and $T_c$, which are robust.
Keeping this caveat in mind, we choose an arbitrary order-one value for $\theta$.
For example, using $\theta = 2$ we find
\begin{align}
 \text{EuS:}\qquad C
 & = 15 \cdot 10^{-3} \, , \qquad \hspace{0.4em}
 f^+
 = 0.77 \, \text{\AA} \, , \\
 \text{EuO:}\qquad C
 & = 5.6 \cdot 10^{-3} \, , \qquad
 f^+
 = 0.72 \, \text{\AA} \, .
\end{align}
The predictions for $C$ are within $<10\%$ error compared to the experimental values in Table \ref{tab:exp-data}, while the estimates for $f^+$ are off by a factor of 2.

\begin{remark} \label{AFremark}
Reference \cite{AharonyFisher} passes from the microscopic theory to the effective theory using an alternative method, which goes back to \cite{Wilson:1971dh,Wilson:1973jj}. Their method amounts to replacing the Heisenberg model integration measure, with its restriction $\vec m^2 =\mu^2$, by a centered Gaussian of width of order $\mu$:
\begin{align}
 \delta(\vec m^2 - 1) \to \exp \left(-\frac{w}{2 \mu^2 } \vec m^2+\ldots \right)  \, ,
 \label{eq:repl-delta}
\end{align}
where $w = 1$ in \cite{AharonyFisher}, but we keep it general to see how it affects the final result. The $\ldots$ includes some quartic interaction terms, which are not important for the present discussion.

After the replacement, the partition function becomes a Gaussian integral:
\begin{align}
	Z
	\to
	\int D\vec B \, \prod_{x} d^3m_x \,
	\exp \left( {-}\beta \Hm[m,B] - \frac{w}{2 \mu^2} \sum_x \vec m_x^2 + \ldots \right) \, .
\end{align}
It is now straightforward to take the continuum limit of the Hamiltonian using \eqref{eq:Ksmooth}, and to integrate out the dynamical $\vec B$ field. In this approach, the continuum limit of $\vec m$ is directly the coarse-grained magnetization, with no additional rescaling needed. The effective Hamiltonian is of the form \eqref{Hpheno}, \eqref{eq:ham-phi}, with effective parameters (AF stands for Aharony-Fisher)
\begin{align}
 a_{\text{AF}}
 = \frac{\mathrm{a}^2 J V}{6} \, , \qquad
 b_{\text{AF}}
 = \frac{w V}{\beta  \mu ^2} - 2 \theta  J V - 4 \pi \, .
\end{align}
Here we also allowed for the ambiguity $\theta$, of the same origin as in Section \ref{sec:HS}. Ref.~\cite{AharonyFisher} does not discuss it and uses directly $\theta = 1/2$.

We now trade the short-range coupling $J$ by the critical temperature $T_c$.
In solving for $J$, we consider $w$ and $\theta$ as number independent of $J$ and $\beta$.
All in all, we find
\begin{align}
 C_{\text{AF}} = \frac{\mu ^2}{w k_B T_c V} \, , \qquad
 f^+_{\text{AF}} = \frac{\text{a}}{2 \sqrt{3 \theta}} \, \sqrt{1-\hat g_{\text{AF}}} \, , \qquad
 \hat g_{\text{AF}}
 = \frac{4 \pi  \mu ^2}{w k_B T_c V} \, .
\end{align}
Comparing these equations to \eqref{eq:ghat} and \eqref{eq:critamp}, we see that the method of \cite{AharonyFisher} does not agree with ours for $w = 1$ used in \cite{AharonyFisher}, while it would agree for $w=3$, up to $O(\hat g)$ corrections in $ f^+$ which are small. Even this partial agreement is quite surprising, given that the replacement in \eqref{eq:repl-delta} is rather ad hoc. It cannot be considered in any sense an \emph{approximation}, as it does not even conserve the rough shape of the potential. Our approach based on the Hubbard-Stratonovich transformation seems better justified.

\end{remark}

\section{Trace of stress tensor in dipolar model}
\label{sec:renorm}

In this appendix we renormalize the dipolar model \eqref{effLagr2}, using methods analogous to \cite{Brown:1979pq}, putting the discussion of sections \ref{sec:stress-tens-arg} and \ref{sec:shift} on solid ground.
To summarize the strategy, we construct a renormalized stress tensor, a renormalized virial current, and a renormalized shift charge.
Furthermore, we show it is possible to improve both the stress tensor and the virial current to make them good scaling operators.
Finally, we show that under shift symmetry the virial current maps to $\phi_i$. By the discussion in Section \ref{sec:shift}, this implies the virial current has dimension $\Delta_V = d-1$.

\subsection{Basic notation}

We work with the Hamiltonian \eqref{effLagr2}, except that we rename $\lambda \to \lambda_0$. Throughout this appendix, $\phi_i$ and $U$ are bare fields and $\lambda_0$ is the bare coupling.
We denote renormalized fields $[\Om]$ and renormalized coupling $\lambda$, namely
\begin{align}
	\phi_i = Z_\phi [\phi_i] \, , \quad
	U = Z_U [U] \, , \quad
	\lambda_0 = \mu^\epsilon \lambda \left(
	1
	+ \frac{a_{11} \lambda + a_{12} \lambda^2 + \ldots}{\epsilon}
	+ \frac{a_{22} \lambda^2 + \ldots }{\epsilon^2} + \ldots
	\right) \, .
\end{align}
The renormalization factors $Z_\Om$ and $a_{ij}$ follow from requiring finiteness of all correlation functions of renormalized operators
\begin{align}
	G_{N,M}(x_1, \ldots, x_N; y_1, \ldots, y_M)
	= \langle
	[\phi_{i_1}](x_1) \ldots [\phi_{i_N}](x_N) \,
	[U](y_1) \ldots [U](y_M) \rangle \, .
\end{align}
From the fact that $\langle \phi_i(x) \, U(y) \rangle$ does not receive perturbative corrections, as discussed below equation \eqref{eq:freeprops}, we conclude that $Z_U = Z_\phi^{-1}$.

The beta function is $\beta(\lambda) = \mu \frac{d\lambda}{d\mu}$ and the anomalous dimensions is $\gamma_\Om = \mu \frac{d \log Z_\Om}{d\mu}$.
Unlike in the main text, we keep track of equations of motion (EOM)
\begin{align}
	E_i
	\equiv \frac{\delta \tilde \Hm}{\delta \phi_i}
	=
	- \partial_j f_{ji}
	+ \partial_i U
	+ \lambda_0 (\phi_k^{2}) \phi_i \, , \qquad
	E
	\equiv \frac{\delta \tilde \Hm}{\delta U}
	= -\partial_i \phi_i \, ,
\end{align}
since they are necessary to show that correlation functions satisfy scaling Ward identities at the fixed point.

The stress tensor, defined by the formula $T^{i j}=-2\frac{\delta \tilde \Hm}{\delta g_{i j}}$ (see note \ref{note:covariant}), reads
\begin{align}
	T_{ij}
	= f_{ik} f_{jk}
	+ \phi_i \partial_j U
	+ \phi_j \partial_i U
	+ \lambda_0 \phi_k^2 \phi_i \phi_j
	- \delta_{ij} \left( \frac{1}{4} f_{kl}^2 + \phi_k \partial_k U
	+ \frac{\lambda_0}{4} \phi^4 \right) \, .
\end{align}
As expected, it is conserved up to EOM:
\begin{align}
	\partial_i T_{ij}
	= - E_i \, \partial_j \phi_i
	- E \, \partial_j U
	+ \partial_i \big( E_i \, \phi_j \big) \, .
	\label{divT}
\end{align}
The trace of the stress tensor works out to be (compare Eq.~\eqref{eq:Tii})
\begin{align}
	T_{ii}
	&= - \frac{\epsilon}{4} \lambda_0 \phi^4
	+ \frac{\epsilon}{2} \, \phi_i E_i
	- \frac{d}{2} \,  U E
	- \partial_i V_i \, , \label{Tiiapp}\\
	V_i
	&= \frac{d}{2} \phi_i U
	+ \frac{\epsilon}{2} \phi_i E
	- \frac{\epsilon}{2} \partial_j
	\left( \frac{1}{2} \phi^2 \delta_{ij} - \phi_i \phi_j \right) \, .
\end{align}
Expressed in terms of bare fields, $T_{i j}$ should be thought of as a bare stress tensor. 
Below we will discuss how to make it finite. The first step is to express $T_{i j}$ and $V_i$ in terms of renormalized couplings and renormalized operators.

\subsection{Composite operators}
\label{sec:composite}

To organize the operators, note that the Hamiltonian \eqref{effLagr2} enjoys a $\mathbb Z_2$ symmetry, under which $\phi_i$ and $U$ are both odd.
There is also a shift symmetry $U(x) \to U(x) + u$ for constant $u$, with associated current $\phi_i$. We discussed this extra symmetry in Section \ref{sec:shift}. The renormalized shift charge is defined by integrating the renormalized current $[\phi_i]$. Thus we have the relation 
\begin{equation}
	Q=Z_\phi [Q] \label{QQ}
\end{equation}
between the bare and renormalized shift charges.

Note that operators neutral under shift symmetry only mix with neutral operators, while operators charged under shift symmetry can mix with everything.
Once all operators that can mix under renormalization are identified, then the precise renormalization factors should be determined from the requirement that correlators with composite operator insertions should be finite.
More precisely, Green's functions of the form
\begin{align}
	G_{N,M}\big(x_1, \ldots, x_N; y_1, \ldots, y_M; [\Om](x)\big)
	= \langle
	[\phi_{i_1}](x_1) \ldots \,
	[U](y_1) \ldots [\Om](x) \rangle \, ,
\end{align}
should be finite.
For EOM terms this criterion immediately shows that they do not renormalize
\begin{align}
	\phi_i E_i = [\phi_i E_i] \, , \qquad
	U E = [U E] \, .
	\label{eq:eomrenorm}
\end{align}
This is because the insertions of $\phi_i E_i $ and of $ U E $ into correlation functions simply generate $\delta$-functions at the positions of $\phi_i$'s and of $U$'s, respectively, as in \cite{Brown:1979pq}, Eq.~(3.10).
A similar argument shows that 
\begin{align}
	\label{eq:phiE}
	\phi_i E = (Z_\phi)^2 [\phi E]\,.
\end{align}

Let us start renormalizing the simplest composite operators, namely $\phi_i^2$ and $\Phi_{ij}:=\phi_i\phi_j - \frac{\delta_{ij}}{d} \phi_k^2$. Note that these are the only scalar and symmetric traceless operators even under $\mathbb Z_2$, neutral under the shift symmetry, and of 4d dimension $\Delta_{4d} = 2$. 
As a result, they can only get multiplicatively renormalized $\phi_i^2 = Z_{\phi^2} [\phi_i^2]$, $\Phi_{ij} = Z_{\Phi} [\Phi_{ij}]$.

The next case of interest is the renormalization of $U \phi_i$.
In this case, mixing occurs with vector operators that are $\mathbb Z_2$-even, either charged or neutral under shift symmetry, and of 4d dimension $\Delta_{4d} = 3$.
A basis of linearly-independent functions is $\{ U \phi_i, \phi_i \partial_j \phi_j, \partial_i \phi^2, \partial_j (\phi_i \phi_j) \}$. So $U \phi_i$ will be a linear combination of the corresponding renormalized operators:
\begin{align}
	U \phi_i
	= \left(1 + \hat c_1\right) \big[U \phi_i \big]
	+{\hat c_2} \, \big[\phi_i E \big]
	+ {\hat c_3} \, \partial_i \big[\phi^2 \big]
	+ {\hat c_4} \, \partial_j \big[\Phi_{ij} \big] \, .
	\label{eq:Uphii}
\end{align}
Here $\hat c_i = \hat c_i(\lambda,\epsilon)$ are counterterms that make the renormalized operators finite, which have an ascending series of poles in $\epsilon$ starting at $O(\epsilon^{-1})$. All such quantities below will carry a hat.
The last two terms are total derivatives, so they only modify the virial current by improvement terms.
We can further argue that $\hat c_1 = 0$. For this let us act by $[Q]$ on both sides of \eqref{eq:Uphii}. Using \eqref{QQ} we obtain:
\begin{equation}
	[\phi_i] = \left(1 + \hat c_1\right) \left[[Q],\big[U \phi_i \big] \right]\,, \label{Qren}
\end{equation}
where we used the fact that all operators but the first in the r.h.s. of \eqref{eq:Uphii} are shift-invariant.
Since the l.h.s. of \eqref{Qren} is finite, the r.h.s. must be finite as well, hence $\hat c_1=0$.

The other case of interest is the renormalization of $\phi^4$. In this case, we need a basis of $\Delta_{4d}=4$ operators, which are shift neutral and $\mathbb Z_2$ even:
\begin{align}
	\{ \phi^4, \phi_i \partial_i U, \phi_i \partial_i \partial_j \phi_j, \phi_i \partial^2 \phi_i, \partial_i (\phi_i \partial_j \phi_j), \partial^2 \phi^2, \partial_i \partial_j (\phi_i \phi_j) \} \, .
	\label{eq:basis4}
\end{align}
Note that one linear combination of operators in \eqref{eq:basis4} corresponds to $\phi_i E_i$, where $E_i$ is the EOM for $\phi_i$.
Similarly, we identify terms $\partial_i \phi_i$ with $E$, the EOM for $U$.
Finally, the term $\phi_i \partial_i U$ is related to the finite operator $U E$ by integration by parts, so it must have a finite integral. As a result, it can only get infinite contributions that are total derivatives:
\begin{align}
	\phi_i \partial_i U
	= [\phi_i \partial_i U]
	+ \hat q_2 \, \partial_i [ \phi_i E]
	+ \hat q_3 \, \partial^2 [\phi^2]
	+ \hat q_4 \, \partial_i \partial_j [\Phi_{ij}] \, .
	\label{eq:renPdU}
\end{align}
Again $\hat q_i = \hat q_i(\lambda,\epsilon)$ are ascending series in poles in $\epsilon$ starting at $O(\epsilon^{-1})$.
Comparing to \eqref{eq:Uphii}, we can rearrange the equation as
\begin{align}
	[\phi_i \partial_i U]
	- [E U]
	- \partial_i [U \phi_i]
	= (\hat c_2 - \hat q_2) \, \partial_i [ \phi_i E]
	+ (\hat c_3 - \hat q_3) \, \partial^2 [\phi^2]
	+ (\hat c_4 - \hat q_4) \, \partial_i \partial_j [\Phi_{ij}] \, .
\end{align}
Since the left-hand side is finite and the right-hand side goes like $O(\epsilon^{-1})$, the right-hand side must vanish.
We conclude that renormalization preserves integration by parts
\begin{align}
	[\phi_i \partial_i U]
	= [E U]
	+ \partial_i [U \phi_i] \, ,
	\label{eq:ibp}
\end{align}
a fact that will be useful below.

After these technical remarks, we see that the most general form of the renormalized quartic field is
\begin{align}
	\frac{\lambda_0}{4} \phi^4
	= \left(1 + \hat k_1\right) \frac{\mu^\epsilon \lambda}{4} \big[\phi^4\big]
	+ \hat k_2 \, [\phi_i  E_i ]
	+ \hat k_3 \, [ \phi_i \partial_i U ]
	+ \hat k_4 \, [ \phi_i \partial_i E ]
	+ \hat k_5 \, \partial_i [ \phi_i E] \nonumber \\
	+ \hat k_6 \, \partial^2 [\phi^2]
	+ \hat k_7 \, \partial_i \partial_j [\Phi_{ij}]  \, .
\end{align}
We will see in a second that $\hat k_1,\ldots,\hat k_4$ only have a simple pole in $\epsilon$, which is a consequence of requiring that $\partial_\lambda G_{N,M}$ is finite.
A similar trick was used in \cite{Brown:1979pq}, Eq.~(3.19).
Using the chain rule, the derivative $\partial_{\lambda_0} G_{N,M}$ inserts an integrated bare operator $\int \phi^4$.
As a result of the integration, all total derivative terms drop out, hence $\hat k_5,\hat k_6,\hat k_7$ cannot be determined by this trick but need an explicit computation of the divergence which we will not do. 

As to the trick, the precise calculation is analogous to \cite{Brown:1979pq}, and one finds\footnote{We point out a minor difference in our notation from \cite{Brown:1979pq}. Ref.~\cite{Brown:1979pq} denotes by $\beta(\lambda)$ the 4d part of the beta function, while the beta function in $d=4-\epsilon$ is given by $\mu \frac d{d\mu}\lambda|_{\rm Brown} =-\epsilon \lambda + \beta(\lambda)$. On the other hand, we denote by $\beta(\lambda)$ the full beta function in $d=4-\epsilon$: $\mu \frac d{d\mu}\lambda|_{\rm here} =\beta(\lambda)$.}
\begin{align}
	\hat k_1 = -\frac{\beta(\lambda) + \epsilon \lambda}{\lambda\, \epsilon} \, , \qquad
	\hat k_2 = -\hat k_3 = \frac{\gamma_\phi}{\epsilon} \, , \qquad
	\hat k_4 = 0 \, ,
\end{align}
which also used integration by parts as in \eqref{eq:ibp}.
All in all, the renormalization of the quartic operator reads
\begin{align}
	\frac{\lambda_0}{4} \phi^4
	= - \frac{\mu^\epsilon \beta(\lambda)}{4 \epsilon} \big[\phi^4\big]
	+ \frac{\gamma_\phi}{\epsilon} \, [ \phi_i E_i ]
	- \frac{\gamma_\phi}{\epsilon} \, [ \phi_i \partial_i U ]
	+ \, \hat k_5, \hat k_6, \hat k_7 \text{ terms}
	\, .
\end{align}
Using this equation and \eqref{eq:Uphii}, we can finally express the trace \eqref{Tiiapp} of the stress tensor in terms of renormalized operators.
The result looks more elegant using a further integration by parts with \eqref{eq:ibp}, giving
\begin{align}
	T_{ii}
	= \frac{\mu^\epsilon \beta(\lambda)}{4} [\phi^4]
	- \left(\Delta_\phi-1 \right) [ \phi_i E_i]
	- \Delta_U [ U E] 
	- \Delta_U \partial_i
	[U \phi_i]-  p\,\partial_i  [ \phi_i E] - \partial_i\partial_j \Om_{ij}\,.
	\label{eq:rentrace}
\end{align}
The scaling dimensions are defined in \eqref{phiU}, the improvement is $\Om_{ij} =a [\phi^2]\delta_{ij}+ b [\Phi_{ij}]$, and furthermore
\begin{gather}
	p = \frac{d}{2}{ \hat c_2}+\frac{\epsilon}2 (Z_\phi)^2 +\epsilon {\hat k_5}\, , \\
	a =  \frac{d}{2}{ \hat c_3}
	-\frac{\epsilon}{2}\left(\frac 12 -\frac 1 d\right) Z_{\phi^2}+ \epsilon{\hat k_6},\qquad
	b = \frac{d}{2}{ \hat c_4} +\frac{\epsilon}{2} Z_{\Phi} + \epsilon{\hat k_7} \,.
\end{gather}
In expressing $p$ we also used \eqref{eq:phiE}. It will follow from the next subsection that $p$ is finite (this is not clear from above).

\subsection{Finiteness of stress tensor}
\label{sec:finite}

Let us discuss the structure of counterterms for the stress tensor and the virial current.
As already mentioned $T_{ij}$ is a bare stress tensor, and it is not in general finite. We see that its trace \eqref{eq:rentrace}, expressed in terms of renormalized fields, involves coefficients $p,a,b$ which are potentially singular as $\epsilon\to 0$.

However, all fields in the divergence of $T_{ij}$ \eqref{divT}, proportional to EOM, are in fact finite (compare \cite{Brown:1979pq}, (3.29)):
\begin{align}
	\partial_i T_{i j} \sim \text{EOM}
	\quad \text{(finite)} \, .
\end{align}
This strongly constrains possible divergences of $T_{i j}$.

Similarly to how we expressed $T_{ii}$ in terms of finite operators, we could do the same for the remaining symmetric traceless part of $T_{i j}$, see e.g. the analysis in \cite{Brown:1979pq} for the $\phi^4$ case. Note that the symmetric traceless and trace parts of $T_{ij}$ do not mix under renormalization, so that analysis would not affect the renormalization of $T_{ii}$ that we already discussed.
To save time, we will avoid renormalizing here the symmetric traceless part of $T_{i j}$ explicitly. However, by now it should not be a surprise that this can be done.

We can then split $T_{i j}$ into a finite piece, which we call $[T_{i j}]$, and a divergent piece that we call $\hat  R_{i j}$:
\begin{align}
	T_{i j}
	= [T_{i j}] + \hat R_{i j} \, .
\end{align}
The requirement that $\partial_i (T_{i j} - [T_{i j}])$ should be finite implies that $\partial_i \hat R_{i j} = 0$, where this conservation does not rely on EOM. To begin with, this implies that the Poincaré charges $P_i$, $M_{ij}$ constructed by integrating $T_{ij}$ and $[T_{ij}]$ coincide. The divergent piece $\hat R_{i j}$ drops out from them - the Poincaré charges are finite.

Furthermore, $\hat R_{i j}$ as any symmetric 2-tensor field satisfying the condition $\partial_i \hat R_{i j} = 0$ can be written in the form:
\begin{align}
	\hat R_{i j} = \partial_k \partial_l \hat Y_{ [i k] [j l]} \, ,
	\label{eq:counter}
\end{align}
where $\hat Y_{[i k] [j l]}$ has symmetries of the Riemann tensor, i.e. is a field antisymmetric in $i k$ and in $j l$ and symmetric under the exchange of these two groups of indices. For $c$-number tensor fields i.e.~smooth mappings $\mathbb{R}^d\to \mathbb{R}$ this follows by using the Poincar\'e lemma twice, together with the (anti)symmetry of the involved fields, see Exercise 5 of Chapter 4 in \cite{Wald}.\footnote{It is also a partial case of the (dualized) generalized Poincaré lemma for mixed-symmetry tensors (see \cite{Dubois-Violette:1999iqe}, Eq.~(7)).} One could worry that perhaps $\hat Y$ is a non-local function of the fields (e.g.~\cite{Wald-worry}, Eq.~(4.7) and below). However, this worry is unfounded. The point is that the Poincar\'e lemma remains valid in the space of local field, a fact known as ``algebraic Poincar\'e lemma'' (see \cite{Barnich:2000zw}, Theorem 4.2). Hence the same argument shows that $\hat Y_{[i k] [j l]}$ is a local field, i.e.~can be built out of products of $U$, $\phi_i$ and their derivatives.

We can check this explicitly for our dipolar model. We can make the most general ansatz for $\hat R_{ij}$ consisting of all rank-two symmetric operators of $4d$ dimension $\Delta_{4d} = 4$ being $\mathbb{Z}_2$ even and shift invariant.
Requiring conservation, we find that $\hat Y_{[ik][jl]}$ is a linear combination of two building blocks
\begin{align}
	(\delta_{il} \delta_{jk} - \delta_{ij} \delta_{kl}) \phi^2 \, , \qquad
	\delta_{il} \Phi_{jk} \pm (\text{3 permutations}) \, ,
	\label{eq:dipol-improv}
\end{align}
and it is indeed local.

An important consequence of equation \eqref{eq:counter} is that all divergent contributions to the bare stress tensor must contain two total derivatives.
Since the $p$ term in \eqref{eq:rentrace} contains only one total derivative, its coefficient $p$ must be finite.

\subsection{Building scaling operators}
\label{sec:scaling}

Up to now, we have explained how to obtain a finite stress tensor and virial current.
We now show how to make them good scaling operators.
For the sake of clarity, in this section we drop square brackets around renormalized operators, e.g. $[T_{ij}] \to T_{ij}$, although all operators are finite.

The argument to make the stress tensor a scaling operator is well known \cite{Polchinski:1987dy},\cite{Nakayama:2013is,Dymarsky:2013pqa}.
One starts from the most general form of the commutation of the dilatation operator $D$ and the stress tensor:
\begin{align}
	[D, T_{ij}]
	= x_m \partial_m T_{ij} + d \, T_{ij}
	+ y_a \partial_k \partial_l Y^a_{ikjl} \, .
\end{align}
Here $Y^a_{ikjl}$ is a complete set of operators with the symmetries of the Riemann tensor (excluding operators such that $\partial_k \partial_l Y_{ikjl}=0$) such that $\partial_k \partial_l Y_{ikjl}$ can mix with the stress tensor. In perturbation theory these are operators of 4d scaling dimension 2.

The operators $Y_{ikjl}$ themselves generically mix under dilatation
\begin{align}
	[D, Y^a_{ikjl}]
	= x_m \partial_m Y^a_{ikjl} + \Delta_{ab} Y^b_{ikjl} \, .
\end{align}
With this information in mind, we can perform a finite improvement\footnote{Ref.~\cite{Polchinski:1987dy} has a mistake in the sign before $\Delta$ in the following equation. The correct sign, here as in \cite{Nakayama:2013is,Dymarsky:2013pqa}, requires changes in the subsequent argument.}
\begin{align}
	T_{ij}
	\to T_{ij}
	+ y^a (d-2-\Delta)_{ab}^{-1} \partial_k \partial_l Y^b_{ikjl} \, ,
	\label{eq:improvT}
\end{align}
such that the new stress tensor is a good scaling operator, or in other words
\begin{align}
	[D, T_{ij}]
	= x_m \partial_m T_{ij} + d \, T_{ij} \, .
	\label{eq:goodTscale}
\end{align}
The only caveat is that improvement \eqref{eq:improvT} is valid provided the matrix $\Delta_{ab}$ does not have any eigenvalue $\Delta=d-2$.
In our theory, there are two candidate improvements \eqref{eq:dipol-improv}. The dimension $\Delta_{\phi^2}$ is known at two loops \cite{DombGreenVol6}, while we computed $\Delta_{\Phi_{ij}}$ at one loop using conformal perturbation theory
\begin{align}
	\Delta_{\phi^2}
	= 2
	- \frac{8 \epsilon }{17}
	+ \frac{1441 \epsilon ^2}{14739}
	+ O(\epsilon^3) \, , \qquad
	\Delta_{\Phi_{ij}}
	= 2
	- \frac{44 \epsilon }{51}
	+ O(\epsilon^2) \, .
\end{align}
Since neither dimension is exactly $d-2$, we can always improve $T_{ij}$ to be a good scaling operator.

Let's imagine we already performed improvement \eqref{eq:improvT}.
Then taking the trace of \eqref{eq:goodTscale}, gives the most general consistent commutation relation  for the virial current\footnote{In more general theories possessing conserved global symmetry currents of dimension $d-1$, those could also appear in the right-hand side. Then, one may not be able to define a good scaling virial current operator. Physically, this would be the effect of mixing between scale transformation and the global symmetry transformation under the RG flow. In our dipolar model, there are no conserved global symmetry currents of dimension $d-1$, the only conserved current being the shift symmetry current $\phi_i$, which cannot appear in the right-hand side because of the $\mathbb{Z}_2$ symmetry and because it does not have the right classical dimension.}
\begin{align}
	[D, V_i] = x_m \partial_m V_i + (d-1) V_i + w_a \partial_j A^a_{ij} \, .
\end{align}
Here $A^a_{ij} = -A^a_{ji}$ is a basis of antisymmetric operators.
As before, the basis behaves under dilations as
\begin{align}
	[D, A^a_{ij}]
	= x_m \partial_m A^a_{ij} + \widehat \Delta_{ab} A^b_{ij} \, .
\end{align}
With this information, we can make $V_i$ a good scaling operator using the freedom to transform $V_i$ in a way that preserves $T_{ii} = -\partial_i V_i$.
The right improvement is
\begin{align}
	V_{i}
	\to V_{i}
	+ w^a (d-2-\widehat\Delta)_{ab}^{-1} \partial_j A^b_{ij} \, .
	\label{eq:improvV}
\end{align}
In the case of the dipolar model, there is no candidate antisymmetric tensor $A^a_{ij}$ with the right dimension, so the virial current improvement is unnecessary, and we do not need to discuss whether $\widehat\Delta_{ab}$ has eigenvalues $d-2$.
However, for other models this discussion might be necessary.

\subsection{Summary}

To wrap up the discussion, sections \ref{sec:composite} and \ref{sec:finite} show that we can find a finite stress tensor $[T_{ij}]$ that generates the Poincaré symmetry charges.
Furthermore, Section \ref{sec:scaling} shows that it is possible to choose suitable improvements such that both $[T_{ij}]$ and $[V_i]$ are scaling operators, of dimensions $d$ and $d-1$.
Combining the results, the trace of the stress tensor is
\begin{align}
	\delta_{ij} [T_{ij}]
	&=\frac{\mu^\epsilon \beta(\lambda)}{4} [\phi^4]
	- \left(\Delta_\phi-1\right) [ \phi_i E_i]
	- \Delta_U [ U E]
	- \partial_i [V_i] \, , \\
	[V_i]
	&= \Delta_U [U \phi_i]
	+ p [ \phi_i E]
	+ q \partial^2 [\phi^2]
	+ r \partial_i \partial_j [\Phi_{ij}] \, .
\end{align}
Recall that $p$, $q$ and $r$ are not determined from our analysis, except they are finite constants.
These operators have all the desired properties discussed in Section \ref{sec:shift}.
Indeed, the trace at the fixed point contains the virial current, and the virial current is mapped to $[\phi_i]$ under shift symmetry:
\begin{align}
	\delta_{ij} [T_{ij}] \big|_{\text{fixed point}}
	= - \partial_i [V_i] + \text{EOM} \, , \qquad
	\big[ [Q], [V_i] \big] \propto [\phi_i] \, .
\end{align}
As explained in Section \ref{sec:shift}, the latter equation is responsible for explaining the ``paradox'' of why the virial current does not acquire anomalous dimension.

\subsection{Scaling Ward identity}

Although it is not strictly necessary for us, before concluding the appendix we derive the Ward identity for scale invariance.
We construct the scale current $D_i = x^j [T_{ij}] + [V_i]$, which satisfies the conservation equation
\begin{align}
	\partial_i D_i
	& = \delta^{ij}[T_{ij}] + \partial_i [V_i] + x^j \partial_i [T_{ij}] \\
	& = \frac{\mu^\epsilon \beta(\lambda)}{4} [\phi^4]
	- E_i \left(\Delta_\phi + x^j \partial_j \right) \phi_i
	- E \left(\Delta_U + x^j \partial_j \right) U
	+ \partial_i \left( E_i \, x^j \phi_j \right) \, .
\end{align}
The last total-derivative term, which was generated by integrating by parts, does not contribute to the Ward identities, which follow from
\begin{align}
	\int d^d x \,
	G_{N,M}\big(x_1, \ldots, x_N; y_1, \ldots, y_M; \partial_i D_i(x)\big)
	= 0 \, .
\end{align}
This is evaluated by recalling that the EOM acts in correlation functions as
\begin{align}
	\Big\langle \frac{\delta \tilde \Hm}{\delta \Om(x)}
	\Om_1(x_1) \ldots \Om_n(x_n) \Big\rangle
	= \sum_{i=1}^n
	\Big\langle \Om_1(x_1) \ldots \frac{\delta\Om_i(x_i)}{\delta \Om(x)}
	\ldots \Om_n(x_n) \Big\rangle \, ,
\end{align}
where $\Om$ is either of $\phi_i$ or $U$.
At the end of the day, we find the expected scaling Ward identity:
\begin{align}
	\left[
	\sum_{a=1}^n \left( x_a^i \frac{\partial}{\partial x_a^i} + \Delta_\phi \right)
	+ \sum_{b=1}^m \left( y_b^i \frac{\partial}{\partial y_b^i} + \Delta_U \! \right)\right] &
	G_{N,M} =
	\frac{\mu^\epsilon \beta(\lambda)}{4} \int d^d x \,
	G_{N,M}\big([\phi^4](x)\big)\, .
\end{align}
For compactness we dropped the arguments $x_a^i$ and $y_b^i$ on the correlators $G_{N,M}$.


\providecommand{\href}[2]{#2}\begingroup\raggedright\endgroup

\end{document}